\documentclass{aa}  

\usepackage{graphicx}
\usepackage{txfonts}
\usepackage{graphicx}
\usepackage{placeins}
\usepackage{float}
\usepackage{subfigure}
\usepackage{amsmath}
\usepackage{mathrsfs}
\usepackage{amssymb}
\usepackage{color,epstopdf}
\usepackage{pdflscape}

\newcommand{\Ygg}{Yggdrasil}
\newcommand{\Odin}{Odin}
\newcommand{\Thor}{Thor}
\newcommand{\Freja}{Freja}
\newcommand{\Loke}{Loke}
\newcommand{\Bifrost}{Bifrost}
\newcommand{\HI}{H{\small I}}

\newcommand{\Lya}{Ly$\alpha$}
\newcommand{\NV}{N{\small V}}
\newcommand{\SiIV}{Si{\small IV}}
\newcommand{\NIV}{N{\small IV}]}
\newcommand{\CIV}{C{\small IV}}
\newcommand{\HeII}{He{\small II}}
\newcommand{\OIII}{O{\small III}]}
\newcommand{\CIII}{C{\small III}]}
\newcommand{\CII}{C{\small II}]}
\newcommand{\CI}{[C{\small I}]}
\newcommand{\CO}{CO(8--7)}
\newcommand{\NHI}{$N_{\text{H{\scriptsize I}}}$}
\newcommand{\NCIV}{$N_{\text{C{\scriptsize IV}}}$}
\newcommand{\dustcont}{235\,GHz}
\newcommand{\HzRG}{H$z$RG}
\newcommand{\HERGE}{HERG{\'E}}
\newcommand{\cmtwo}{cm$^{-2}$}

\begin{document} 

   \title{The Mysterious  Morphology of MRC0943-242 \\ as Revealed by ALMA and MUSE}


   \author{Bitten Gullberg \inst{1}
    \and
    Carlos De Breuck\inst{1}
    \and
    Matthew D. Lehnert \inst{2}
    \and
    Jo\"{e}l Vernet\inst{1}
    \and
    Roland Bacon\inst{3}
    \and
    Guillaume Drouart \inst{4}
    \and
    Bjorn Emonts \inst{5}
    \and
    Audrey Galametz \inst{6}
    \and 
    Rob Ivison \inst{1,7}
    \and
    Nicole P. H. Nesvadba \inst{8}
    \and
    Johan Richard \inst{3}
    \and
    Nick Seymour \inst{9}
    \and
    Daniel Stern \inst{10}
    \and
    Dominika Wylezalek \inst{11}
              }

   \institute{European Southern Observatory, 
              Karl-Schwarzschild-Str. 2, D-85748 Garching\\
              \email{bgullber@eso.org}
    \and
    Institut d'Astrophysique de Paris, UMR 7095, CNRS, Universit\'e Pierre et Marie Curie, 98bis boulevard Arago, 75014, Paris, France 
    \and
    CRAL, Observatoire de Lyon, CNRS, Universit\'e Lyon 1, 9 avenue Ch. Andr\'e, 69561, Saint Genis-Laval Cedex, France
    \and
    Department of Earth and Space Science, Chalmers University of Technology, Onsala Space Observatory, 43992, Onsala, Sweden
    \and
    Centro de Astrobiolog\'ia (INTA-CSIC), Ctra de Torrej\'on a Ajalvir, km 4, 28850 Torrej\'n de Ardoz, Madrid, Spain 
    \and
    Max-Planck-Institut f\"ur Extraterrestrische Physik, Giessenbachstra\ss e 1, 85748 Garching, Germany
    \and
    Institute for Astronomy, The University of Edinburgh, Royal Observatory, Blackford Hill, Edinburgh EH9 3HJ, UK
    \and
    Institut d'Astrophysique Spatiale, CNRS, Universit\'e Paris-Sud, Bat. 120-121, F-91405 Orsay, France
    \and
    International Centre for Radio Astronomy Research, Curtin University, Perth WA 6845, Australia
    \and
    Jet Propulsion Laboratory, California Institute of Technology, Pasadena, CA 91109, USA
    \and
    Department of Physics and Astronomy, Johns Hopkins University, 3400 N. Charles St, Baltimore, MD, 21218, USA
             }

   \date{\today}

 
\abstract{We present a pilot study of the $z=2.923$ radio galaxy
MRC0943-242, where we for the first time combine information from ALMA
and MUSE data cubes.  Even with modest integration times, we disentangle
an AGN and a starburst dominated set of components. 
These data reveal a highly complex morphology, as the AGN,
starburst, and molecular gas components show up as widely separated
sources in dust continuum, optical continuum and CO line emission
observations.  CO(1--0) and CO(8--7) line emission suggest that there is a molecular
gas reservoir offset from both the dust and the optical continuum that is
located $\sim$90\,kpc from the AGN. The UV line emission has a complex
structure in emission and absorption. The line emission is mostly due to
\textit{i}) a large scale ionisation cone energised by the AGN, \textit{ii}) a \Lya\ emitting
bridge of gas between the radio galaxy and a heavily
star-forming set of components.  Strangely, the ionisation cone
has no \Lya\ emission.  We find this is due to an optically thick layer
of neutral gas with unity covering fraction spread out over a region of at least $\sim100$\,kpc from the AGN. 
Other, less thick absorption components are
associated with \Lya\ emitting gas within a few tens of kpc from the radio
galaxy and are connected by a bridge of emission. We speculate that
this linear structure of dust, \Lya\ and CO emission, and the redshifted
absorption seen in the circum-nuclear region may represent an accretion
flow feeding gas into this massive AGN host galaxy.}


   \maketitle
%

\section{Introduction}

Powerful high-$z$ radio galaxies (H$z$RGs), defined as having $L_{3\,GHz}
> 10^{26}$\,W\,Hz$^{-1}$  and $z > 1$, are unique markers of the most
active galaxies in the early Universe, showing signatures of both
luminous AGN activity and vigorous starbursts.  \HzRG s are extremely
luminous in both the mid-IR \citep{ogle06, seymour07, debreuck10} and the
sub-mm waveband \citep{archibald01,reuland04,drouart14}. This has been
interpreted as evidence of high black hole accretion rates, combined
with high star-formation rates (SFRs). \HzRG s are some of the most
massive galaxies known at any redshift, with $M> 10^{11}\,$M$_{\odot}$
of stars \citep{seymour07, debreuck10}, confirming prior indications
of their large masses from the tight correlation of the observed
near-IR Hubble K-$z$ diagram for these sources \citep{lilly84, eales97,
debreuck02}. In order to be so luminous, the black holes in these galaxies
must be massive.  Given that stellar bulge mass correlates with black
hole mass \citep[e.g.][]{tremaine02}, it is therefore no surprise that
the most powerful radio galaxies reside in the most massive stellar hosts.

Disentangling the AGN and starburst components requires excellent sampling
of the spectral energy distribution (SED).  \cite{seymour12} illustrate
the decomposition of the starburst and AGN components using 3.5 --
850\,$\mu$m photometry in the $z = 2.16$ radio galaxy, PKS 1138-262,
showing that both have roughly equal contributions to the IR luminosity.
Disentangling the SED of \HzRG \ is the main goal of our HErschel Radio
Galaxy Evolution (\HERGE) project \citep{drouart14}.  The 71 \HzRG\
targeted in the \HERGE \ project are uniformly distributed across $1.0 <
z < 5.2$ with a range of radio powers.  \textit{Spitzer, Herschel}, SCUBA
and LABOCA data suggest that the sources targeted in the HERG{\'E} project
have very high far-IR (FIR) luminosities ($\sim10^{13}\,$L$_{\odot}$,
\citealt{drouart14}), but the data do not have high enough spatial
resolution to pinpoint the AGN host galaxies as the source of the FIR
emission. Resolution is important, as spectral decomposition alone is
probably insufficient to understand the evolutionary state of \HzRG s.

Previous single sources studies of \HzRG s have revealed
that these systems have highly complicated morphologies, e.g. with
line emission separated from the continuum emission.
In several systems (e.g. 4C41.17 and 4C60.07, TXS0828+193, and B3 J2330+3927), most of the gas and dust emission originates
from the companion rather than the AGN host galaxy, while others
(e.g. MRC0114-211, MRC0156-252, and MRC2048-272) show no evidence for a companion \citep{debreuck03,debreuck05,ivison08,ivison12,nesvadba09,emonts14}.  These observations
confirm that spectral decomposition alone is insufficient to understand
the evolutionary state of \HzRG s, sufficiently high spatial resolution
to separate out individual emission components is also crucial.

The sub-arcsecond spatial resolution of ALMA will undoubtedly reveal
a much more complex composition of AGN and starburst dominated sources
than previously thought.  To illustrate this point, we conducted a pilot
study of MRC0943-242 at $z=2.9$, where we combine - for the first time -
sensitive ALMA 235\,GHz and VLT/MUSE observations.  This combination
allows us to trace the spatially-resolved distributions of the
ionisation state of the warm ionised medium (T$\sim$10$^4$\,K) through
the UV emission and absorption lines (e.g. CIV), the warm neutral gas
(T$\sim$ few $\times10^3$\,K) through \Lya\ absorption, and the dense and
diffuse molecular gas as probed by the dust and high-$J$ CO transitions.
Through this combination it is possible to get a comprehensive
picture of several important phases of gas in the interstellar medium
and haloes of \HzRG s.

MRC0943-242 is an ultra steep spectrum radio source with a spectral index of
$\alpha=-1.5$ between $1.4-30$\,GHz and shows no evidence of a spectral
curvature within this range \citep{carilli97, emonts11}.  MRC0943-242 is
located in a proto-cluster \citep[with an over density of almost $5\sigma$;][]{wylezalek13} 
surrounded by many nearby companions detected
in \Lya\ with known redshifts \citep{venemans07} and a giant quiescent
\Lya\ halo with a diameter of $\geq100$\,kpc - extending far beyond the
radio structure \citep{villar-martin03}.  A deep absorption trough in the
\Lya\ emission line suggests the presence of a large amount of neutral
gas \citep{roettgering95}.  Based on a 1.5\,\AA-resolution spectrum
\cite{roettgering95} conclude that the saturated absorption trough is
due to a reservoir of neutral hydrogen (\HI) situated within the \HzRG \
it self. They determine a \HI\ column density of \NHI\ $=10^{19}$\,\cmtwo\
in the absorber.  Using the same observational set-up, \cite{binette00}
determine a \HI\ column density similar to that of \cite{roettgering95}
and a \CIV \ column density of \NCIV\ $=10^{14.5\pm0.1}$\,\cmtwo. However
based on Voigt-profile fittings to the absorption troughs in both the \Lya
\ and \CIV \ emission lines \cite{binette00} conclude that the reservoir
of gas causing the absorption features is not situated within the \HzRG,
but rather due to a low metallicity ($Z=0.01$\,Z$_{\odot}$), low density
gas in the outer halo at a redshift of $z=2.29202\pm0.0002$. They propose
that this gas is a remnant of gas expelled from the \HzRG \ during
the initial starburst. Though such a starburst is expected to enrich
the expelled gas more than what is observed for MRC0943-242,
\cite{jarvis03} still favored this scenario. Based on numerical
simulations by \cite{bekki98}, which predict that the outer regions of
galactic halos should be less metal-rich, \cite{jarvis03},
as \citealt{binette00}, favored a picture where the shell lies in
the halo to explain its the low metallicity. Furthermore, by comparing
MRC0943-242 with another \HzRG \ (MRC0200+015), they argue that the age
of the system can have an influence on the metallicity of the shells.
They conclude that the metallicities of absorption shells surrounding
\HzRG \ vary from galaxy to galaxy and from shell to shell. This implies
that the shells must be enriched by a variety of processes.  With
UVES data with a spectral resolution of 25000-40000 \cite{jarvis03}
identified three additional weaker absorbing components of different \HI\
column densities.

In this paper, we present a study of MRC0943-242, where we for the
first time combine relatively short ALMA sub-millimetre (submm)
observations with MUSE optical observations. This ALMA-MUSE-pilot
study reveals an even more complex morphology of MRC0943-242 than
previously seen with star-formation taking place far outside the AGN
host galaxy.  In \S~\ref{sec:obs} we present our ALMA submm and MUSE
optical observations. The results of these observations are given in
\S~\ref{sec:res} and are analysed and discussed in \S~\ref{sec:disc}. In
\S~\ref{sec:speculation} we speculate as to the origin and role of the
substructures and in \S~\ref{sec:con} we summarize our conclusion.

\section{Observations}\label{sec:obs}

\subsection{ALMA observations}

The Atacama Large Milimeter/submilimeter Array (ALMA) cycle 1
observations in Band 6 were carried out on 2014 April 29 for 3\,min
on-source time with 36 working antennas.  The four 1.875\,GHz spectral
windows were tuned to cover the frequency ranges $233.6-237.1$\,GHz and
$248.6-252.3$\,GHz.  We used the Common Astronomy Software Applications
(CASA) and the supplied calibration script to calibrate the data, produce
the data cube and moment 0 maps. The average $T_\text{sys}$
was $\sim$80\,K and the average precipitable water vapour 1.3\,mm.
The quasar J1037-2934 was used as bandpass calibrator and results in
an astrometric accuracy of 4\,mas. The UV coverage was sparse, but uniform 
over $\sim 400$\,k$\lambda$.  In order to search for CO(8--7)
line emission at $\nu_{\text{rest}}=921.80$\,GHz, which is
within the frequency range at $z=2.923$, we subtracted the continuum in
the UV-plane by fitting a first-order polynomial to all four spectral windows, as no strong line emission is contaminating the continuum. 
As our data are limited by the signal-to-noise ratio (S/N), we use a natural weighting (robust parameter of 2). This results in a $0\farcs7\times0\farcs6$
beam with pa~$ =75^{\circ}$ and an RMS of 0.1\,mJy for the continuum.
We binned the line data to 10\,km/s which has an RMS of 1.7\,mJy.
As objects are detected up to 12\arcsec\ from the phase centre, we applied a primary beam correction on all our maps and spectra.

\subsection{MUSE observations}

Multi Unit Spectroscopic Explorer (MUSE) \citep{bacon10} observations were
obtained on 2014 February 21, during the first commissioning run of the
instrument \citep[see][]{bacon14} on VLT/UT4. This one hour observation
taken under $\sim1\arcsec$ seeing was split into three 20\,min exposures
taken at 45, 135 and 225 degree position angle with a small dithering
offset to minimise the effect of systematics. In order to observe the
\Lya\ line of MRC0943-242 at 476.5\,nm, we used the extended wavelength
range mode (i.e. without second order blocking filter), 
resulting in a observed
wavelength range of 460--935\,nm).  These data were taken specifically
to compare MUSE observations with previous observations with other
instruments of this well-studied source.  Thus it serves as a pilot
study for using MUSE to study \HzRG s. All data were processed with the
version 1.0 of the MUSE pipeline \citep{weilbacher12} to produce a fully
calibrated and sky subtracted data cube.  Finally, the cube was cleaned
for sky subtraction residuals with the Principal Component Analysis (PCA)
based algorithm ZAP developed by Soto et al (in prep.).  To analyze the
line emission and to remove a few artefacts in the data, we continuum
subtracted the cubes around specific emission lines such as \Lya.

\begin{figure*}
\centering
\includegraphics[trim=0.65cm 0cm 6.1cm 0.9cm, clip=true,scale=0.67,angle=90]{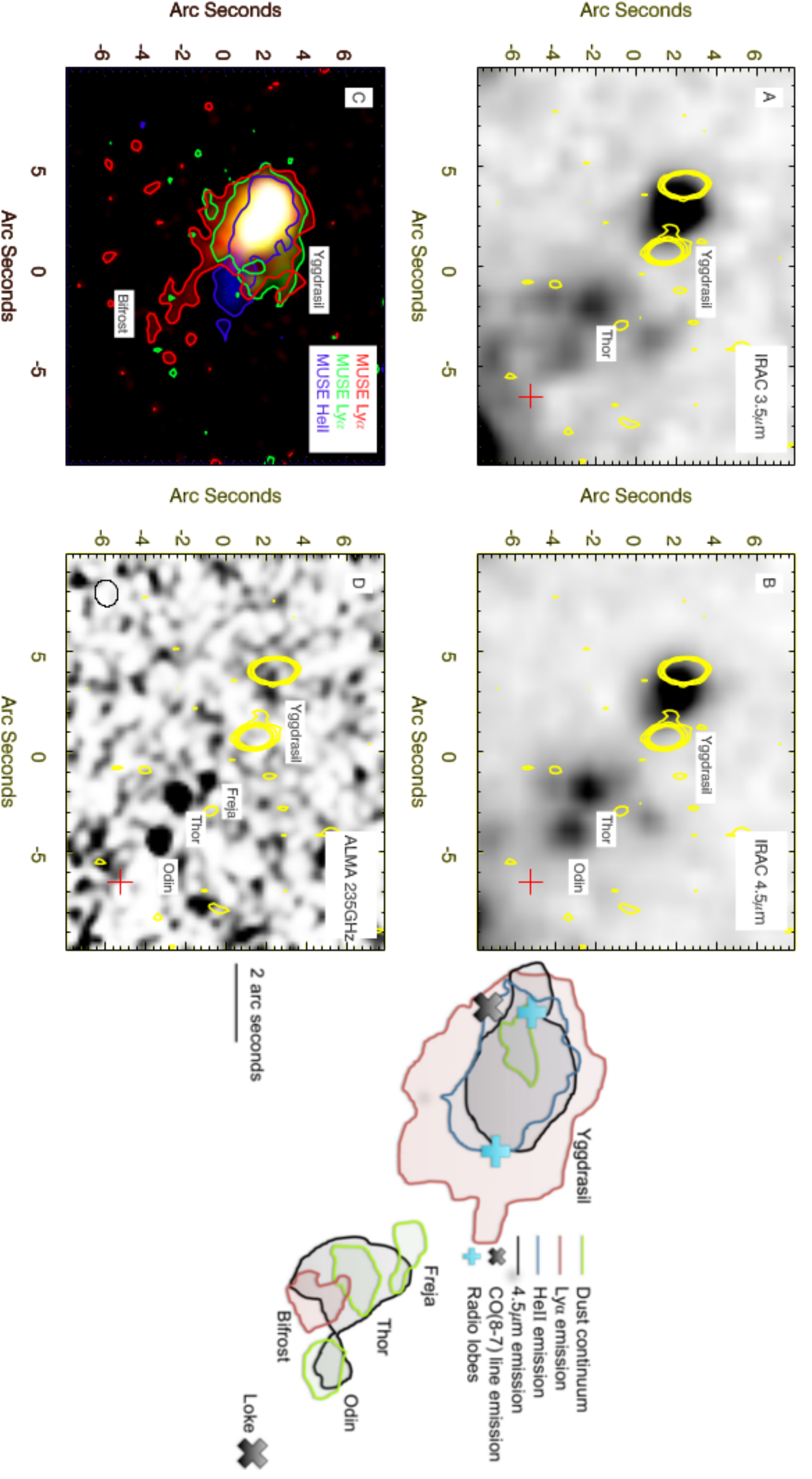}
\caption{{\small Overview of the IRAC (3.6$\,\mu$m and 4.5$\,\mu$m),
MUSE (\Lya\ and \HeII) and ALMA (\dustcont) maps.  \textit{Panel A
and B}: The IRAC 3.6$\,\mu$m and 4.5$\,\mu$m images showing both
\Ygg\ and \Thor. Both IRAC images have a spatial sampling
of 0.61\arcsec/pixel. 
\textit{Panel C}:
A red (\Lya), green (\Lya) and blue (\HeII) image composed of
moment-0 maps of the MUSE cube, which have an spatial
sampling of 0.2\,\arcsec/pixel. The red \Lya\ moment-0 is summed
over $\lambda_{\text{obs}}=4768.8-4776.2$\,\AA (see red bar in panel
A of Fig.~\ref{fig:RGB}), the green \Lya\ moment-0 is summed over
$\lambda_{\text{obs}}=4754.3-4758.8$\,\AA\ (see green bar in panel
A of Fig.~\ref{fig:RGB}) and the blue \HeII\ moment-0 map is summed
over $\lambda_{\text{obs}}=6422.5-6430.0$\,\AA. The red \Lya\ reveals a
bridge of emission connecting \Ygg\ and \Bifrost\ (see \S~\ref{sec:res}), while the green \Lya\
emission shows extended emission to the west. The Blue \HeII\ shows
an extended tail of \HeII\ emission towards the WSW, which is not seen
in \Lya\ emission (see \S~\ref{sec:ion}).  \textit{Panel D}: The ALMA
dust continuum map reveals weak dust emission at the position of the
AGN, but strong dust emission in three aligned components 48-65\,kpc
SW of the AGN. The mm continuum flux density of all four continuum sources
are extracted with apertures of; \Ygg: $2.6\arcsec\times1.7\arcsec$,
\Freja: $1.8\arcsec\times1\arcsec$, \Thor; $1.9\arcsec\times1.5\arcsec$
and \Odin: $1.2\arcsec\times1.4\arcsec$. \textit{Far right}: Schematic
overview of the multi-wavelength components detected in MRC0943-242.
VLA 4.5\,GHz radio observations have been overlaid in yellow contours in Panel A, B and D. The plotted contour levels are for $-3\sigma$, $2\times3\sigma$, $3\sqrt{2}\times3\sigma$, $5\sqrt{2}\times3\sigma$, which is the same for all VLA contour levels through out the paper. The position of \Loke\ is marked with a red cross in Panel A, B and D.}}
\label{fig:morph}
\end{figure*}

\subsection{Previous Supporting \textit{Spitzer} Observations}

MRC0943-242\ was observed among 68 other \HzRG\ in the redshift
range $1<z<5.2$ in a large \textit{Spitzer} survey \citep{seymour07}.
MRC0943-242\ was observed with the Infrared Array Camera (IRAC),
the Infrared Spectrograph (IRS) and the Multiband Imaging Photometer
(MIPS) on \textit{Spitzer}, which cover the wavelengths $3.6\,\mu$m,
$4.5\,\mu$m, $5.8\,\mu$m, $8.0\,\mu$m, $16\,\mu$m, $24\,\mu$m, $70\,\mu$m
and $160\,\mu$m (see \citealt{seymour07} for details on the observations
and reductions).  At 3.6 and 4.5\,$\mu$m, we use the deeper IRAC observations (Fig.~\ref{fig:morph}) obtained as part of the Clusters Around Radio Loud AGN project \citep[CARLA;][]{wylezalek13}.The IRAC  and MIPS observations reveal that
the continuum emission associated with the dust features in our ALMA
data is very diffuse and faint (see section \S~\ref{sec:stellarmass}
for more details). In support of the \textit{Spitzer} data,
we also examined the observations of MRC0943-242 from Keck and
\textit{HST/NICMOS} \citep{vanBreugel98,pentericci01}.  We do
not see any evidence at sub-arcsecond resolution of multiple components.

\section{Results}\label{sec:res}

Combining for the first time ALMA submm observations with MUSE optical
observations reveals an even more complex morphology than seen previously:
with multiple components where only the AGN dominated component is
visible in all wavelengths.  Figure~\ref{fig:morph} shows the IRAC
3.6\,$\mu$m and 4.5\,$\mu$m image, \Lya\ and \HeII\ moment-0 maps in a
colour image observed with MUSE and the \dustcont\ map observed with ALMA.
The \Lya\ moment-0 map shows a bright source at the position of the \HzRG\
(hereafter \Ygg) and an additional component to the South-West (SW),
connected by a bridge of \Lya\ emission (hereafter \Bifrost). \Ygg\
is visible in all four images, while \Bifrost\ splits up into three
components in the \dustcont\ observations (hereafter \Freja, \Thor\
and \Odin). \Thor\ is visible in the IRAC 3.6\,$\mu$m and 4.5\,$\mu$m
images with an extension towards both \Freja\ and \Odin. A \CO\ emitting
component (hereafter \Loke) is located even further to the SW than \Freja,
\Thor\ and \Odin.  The complex morphology is illustrated by the schematic
overview in Fig.~\ref{fig:morph} and additionally shows the
positions of the radio lobes, and the \CO\ emission at the position of
\Ygg\ and \Loke. We now discuss each of these phases in detail.

\begin{figure*}
\includegraphics[trim=0cm 0.7cm 9.1cm 0.9cm, clip=true, scale=0.7, angle=90]{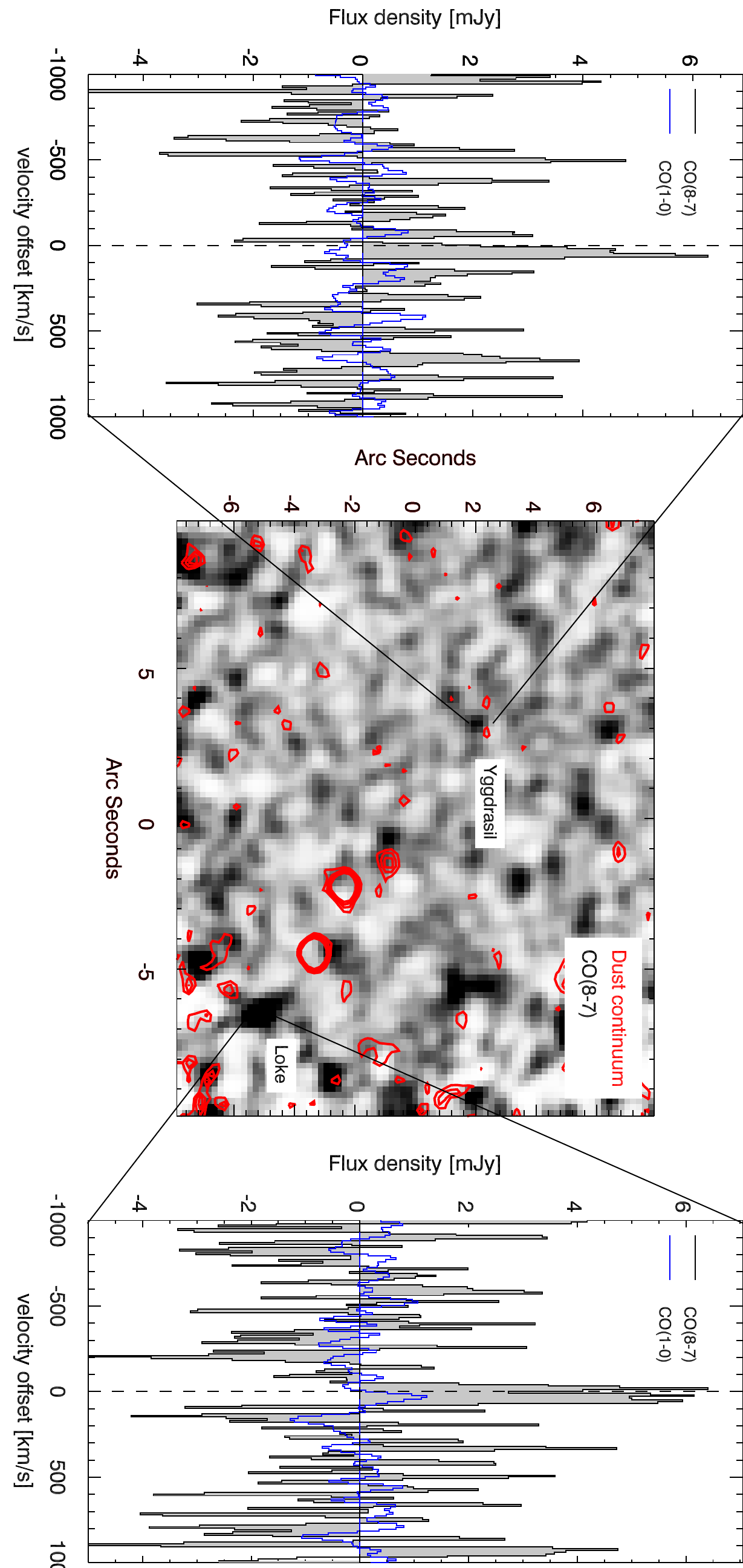}
\caption{{\small The ALMA cube shows CO(8--7) emission at two positions
in the data cube: at the location of Yggdrasil and to the SW in an
isolated component: Loke.  \textit{Middle panel:} The grey scale image
of the CO(8--7) emission overlaid with the ALMA dust continuum contours.
Note the increase in noise towards the edges due to the primary beam correction.
\textit{Left and Right:} The CO(8--7) velocity profiles for both Yggdrasil
(left) and Loke (rigth). Surprisingly the CO(8--7) lines only have a
small offset from the \HeII \ systemic redshift. The spectra
are extracted from a beam-sized area at the position of the emission,
and have RMSs of 1.5\,mJy and 1.0\,mJy respectively. Both detections
are unresolved at the S/N of our data.}}
\label{fig:CO87}
\end{figure*}

\subsection{Surprising dust and molecular gas distribution}

\subsubsection{Dust continuum emission}

\Ygg \ shows up in the ALMA submm observations as a weak dust continuum
emission source at \dustcont, with an extracted flux density over the
region of $0.8\pm0.2$\,mJy (see Table~\ref{table:data}).  However,
the vast majority of the 235\,GHz dust emission originates from three
components (\Freja, \Thor\ and \Odin) aligned along a ``string'' shifted
$48-65$\,kpc towards the SW relative to the position of Yggdrasil (see
Fig~\ref{fig:morph}). Table~\ref{table:data} lists their flux densities
extracted using CASA.

Considering the number counts of $>0.5$\,mJy sources at 235\,GHz
of $\sim2\times10^4$\,deg$^{-2}$ \citep{laurent05}, and the size of the area
within the 8\farcs7 radius from \Ygg\ to \Odin, we expect to find 0.37 sources with $F_{\dustcont}>0.5$\,mJy. We detect four sources within this area; the probability
that these four sources are unrelated is therefore 0.37$^4$, or 2\%. The
probability that these four sources are associated to MRC0943-242 is
thus 98\%.

\subsubsection{Molecular emission lines}

Searching the ALMA cube for molecular gas tracers, we find \CO\
emission at two different positions. One at the position of \Ygg\ and one $\sim90$\,kpc SW
of \Ygg.  The $3.5\sigma$ \CO\ detection at the position of \Ygg\ (see
Fig~\ref{fig:CO87}) is only shifted by $\sim+43$\,km/s from the \HeII\
systemic redshift. Fitting a single Gaussian to this narrow velocity
profile yields a FWHM of $47\pm13$\,km/s, and a velocity integrated
line flux of $0.33\pm0.06$\,Jy\,km/s.  The other 3.5$\sigma$
\CO\ detection is located $\sim90$\,kpc SW of \Ygg\ (\Loke, see
Fig.~\ref{fig:morph} and \ref{fig:CO87}).  Surprisingly, \Loke\ is not
detected in any dust continuum nor optical counter parts.  Like the \CO\
line detected at the position of \Ygg, this \CO\ line is narrow with a
FWHM of $53\pm17$\,km/s, and only shifted by $\sim+16$\,km/s from the
\HeII\ systemic redshift (see Fig.\ref{fig:veldist}). The RMS
of the \CO\ spectrum for \Ygg\ is 1.8\,mJy and 2.1\,mJy for \Loke. This
small difference is due to a combination of the primary beam correction (affecting only \Loke) and continuum subtraction residuals (affecting only \Ygg, as no continuum is detected in \Loke).

To verify the reality of the \CO\ emission, ensuring that these \CO\ lines
are not rare but insignificant noise peaks in the cube, we create 13 moment-0
maps at different frequencies in the cube by collapsing a velocity ranges
in the cube of 180\,km/s. This corresponds to three times the average FWHM of two \CO\ lines and is the same velocity width as for the
\CO-moment-0 map (see Fig.~\ref{fig:CO87}). From these, we select the
brightest areas in each moment-0 map and extract the spectrum from the
cube for each of these areas and ensure that there is no overlap between
velocity regions. From this we find 81 unique peaks for which we extract over the full $233.5-236.5$\,GHz frequency range.
We fit a Gaussian profile to the brightest
peak in each spectrum and determine the S/N in each spectrum. Only two
spectra show a peak of intensity with a S/N of $\sim3$ and a FWHM
of $\sim70$\,km/s which could resemble a real line. The chance
of detecting a line such as \Loke\ within 27$\pm$6\,km/s from \Ygg\
within our 4000\,km/s bandwidth is $27/4000=0.67\%$. In addition, the
probability of finding a $\sim 50$\,km/s wide line like \Ygg\ and \Loke\
is $50/4000=1.25\%$. The combined probability of finding the observed \CO\
configuration in \Loke\ and \Ygg\ is therefore only 0.84\%.

Searching the archival Australian Telescope Compact Array (ATCA) data for
\Ygg\ (see \citealt{emonts11} for more details about the data
reduction and data characteristics) by re-shifting the velocities to
the redshift ($z=2.92296\pm0.00001$ see \S~\ref{sec:ion})
used for this work, we find tentative CO(1--0) line
emission at a 2.8$\sigma$ level at the position of \Loke\ (see
Fig.~\ref{fig:CO87}), consistent with the existence of a large molecular
gas reservoir. The tentative detection of CO(1--0) emission and the CO(8--7)
to CO(1--0) line flux ratio of $\sim0.03$ would imply the presence
of a low excitation molecular gas reservoir.  No CO(1--0) emission was
found at the position of \Ygg\ (see Fig.~\ref{fig:CO87}). This agrees
with the expectation that we should find more highly excited gas near
a powerful radio-loud AGN \citep[e.g.][]{weiss07}

Unfortunately the ATCA observations are limited by the large beam
and low S/N, and it is therefore not possible to determine whether
or not CO(1--0) emission originates from the dust continuum sources,
a free molecular gas component in the halo or both.  In an attempt to
explore if an extended \CO\ component is present, we applied a larger
beam to the ALMA \dustcont\ data. Unfortunately, tapering the ALMA 235
GHz data and convolving them with a beam comparable to that of ATCA
did not produce useful results, as our shallow ALMA observation do not
contain sufficient short baselines. We plan JVLA CO(1--0) and ALMA \CI\
observations at arc-second spatial resolution to pinpoint the exact
location and determine the characteristics of the molecular gas reservoir.

No \CO\ line emission is observed at the positions of \Freja,
\Thor\ and \Odin. We therefore take the $3\sigma$ upper limit of the
line emission at these positions to be 3 times the RMS of the spectra
extracted with the same region sizes as for the continuum emission
(see Table~\ref{table:fluxes}).

\begin{figure*}
\centering
\includegraphics[trim=0.1cm 0.3cm 0cm 0.7cm, clip=true, angle=90, scale=0.675]{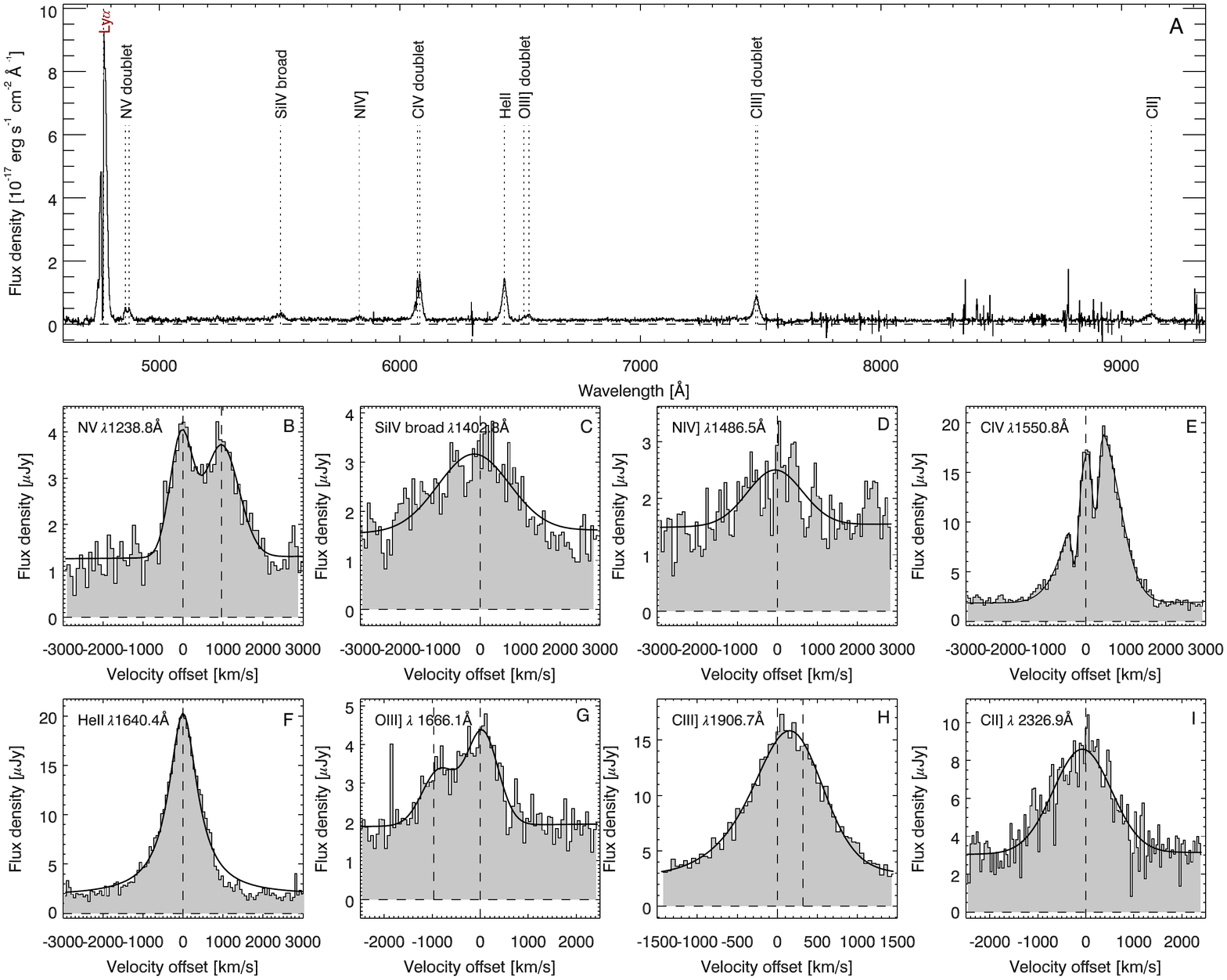}
\caption{{\small The top panel shows the full MUSE spectrum for \Ygg. All
detected lines (other than \Lya) are indicated with a black dotted lines, and the
velocity profiles are shown in Panel B-I. The velocity profiles for \NV,
\SiIV, \NIV, \CIV, \HeII, \OIII,\CIII,\ and \CII\ assume \HeII\ as the
systemic redshift. The \CIV\ emission line shows absorption troughs like
those of the \Lya\ emission line.}}
\label{fig:ions}
\end{figure*}

\begin{table*}      
\centering          
\begin{tabular}{l l l c c }
\hline\hline  
\multicolumn{5}{c}{MUSE} \\
\hline
Line & $\lambda_{\text{rest}}$& $\lambda_{\text{obs}}$ & Line flux & FWHM \\
        & \AA                              & \AA                                & $10^{-16}$\,erg/\cmtwo/s &    km/s  \\
\hline
\multicolumn{5}{l}{\textbf{\Ygg}} \\
\Lya  & $1215.7$ & $4768.9\pm0.2$  & $26.60\pm0.25$  & $1592\pm44$ \\
\NV   & $1238.8, 1242.8$ & $4859.1\pm0.6, 4875.5\pm0.9$  & $0.90\pm0.04$     & $1803\pm174$\\
\SiIV & $1402.8$ & $5500.1\pm1.1$   & $0.60\pm0.04$       & $2493\pm214$\\
\NIV  &  $1486.5$ & $5830.3\pm1.9$  & $0.27\pm0.04$     & $1764\pm370$\\
\CIV  & $1548.2, 1550.8$ & $6073.9\pm1.5, 6072.2\pm1.4$ & $4.34\pm0.06$    & $1410\pm188$\\
\HeII & $1640.4$  & $6435.2\pm0.2$   & $3.04\pm0.05$   & $862\pm30$ \\
\OIII  & $1660.8, 1666.1$ & $6516.3\pm1.4, 6533.8\pm2.5$    & $0.50\pm0.03$     & $1492\pm292$\\
\CIII  & $1906.7, 1908.7$ & $7479.2\pm0.9, 7490.9\pm3.1$     &    $2.00\pm0.06$   & $1734\pm124$\\
\CII   & $2326.0$  & $9122.4\pm1.0$   & $0.90\pm0.04$ & $1606\pm118$\\
\hline
\multicolumn{5}{l}{\textbf{\Bifrost}} \\
\Lya   & $1215.7$ & $4769.2\pm1.1$     &  $0.65\pm0.06$    & $1018\pm205$\\
\hline\hline
\multicolumn{5}{c}{ALMA} \\
\hline
Line & $\nu_{\text{rest}}$& $\nu_{\text{obs}}$ & $SdV$ & FWHM \\
        & GHz                      & GHz                   & Jy\,km/s &    km/s  \\
\hline
\multicolumn{5}{l}{\textbf{\Ygg}} \\
\CO  & 921.8 & $234.94\pm0.1$ &$0.33\pm0.07$ & $43\pm13$ \\
\multicolumn{5}{l}{\textbf{\Loke}} \\
CO(1--0) & 115.3 &  ---  &$<0.08$ & $27\pm16$\\ 
\CO   & 921.8 & $234.96\pm0.1$ &$0.54\pm0.10$ & $57\pm17$ \\
\multicolumn{5}{l}{\textbf{\Freja, \Thor\ and \Odin}} \\
\CO   & 921.8 & --- & $<0.13$& --- \\ 
\CO   & 921.8 & --- & $<0.18$& --- \\
\CO   & 921.8 & --- & $<0.16$& --- \\
\hline
\end{tabular}
\caption{{\small Velocity integrated fluxes and FWHM of the resonance
, fine-structure, and molecular lines for \Ygg, in-between \Odin \
and \Thor \ and \Loke. The ATCA observations have a beam size
$11.5\arcsec\times9.0\arcsec$ and PA $87.5^\circ$. The $3\sigma$ upper
limits on the \CO\ line emission for \Freja, \Thor\ and \Odin\ are given assuming a FWHM of 50\,km/s, i.e. similar to the \CO\ lines for \Ygg\ and \Loke.}}
\label{table:fluxes}
\end{table*}

\subsection{Ionised gas}\label{sec:ion}
\begin{figure}[t]
\includegraphics[trim=2.1cm 2.6cm 2.9cm 5.2cm, clip=true, scale=0.45, angle=90]{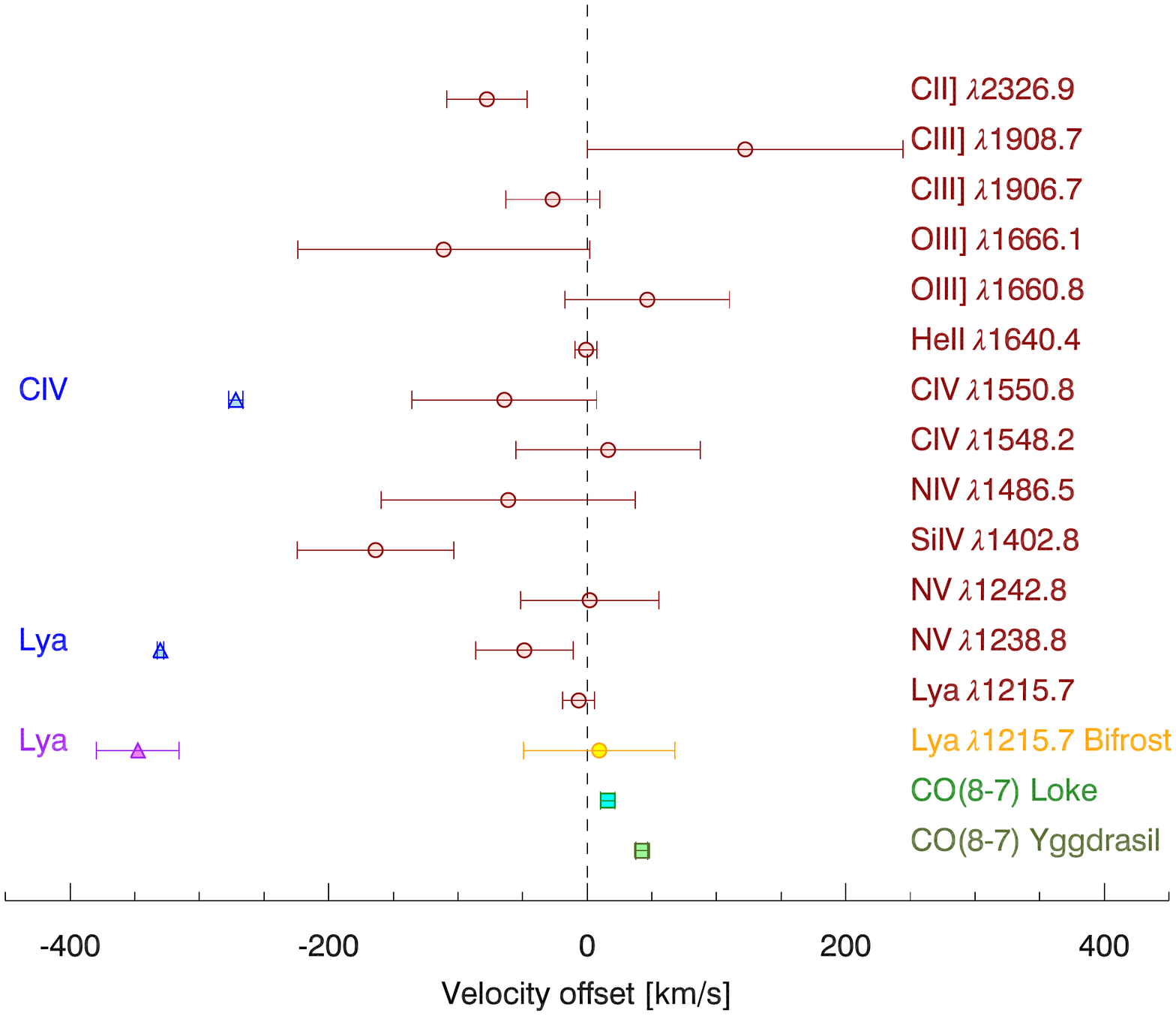}
\caption{{\small The offset of the emission line and absorption features
centers from the \HeII \ systemic redshift.  Red points are emission lines
observed for \Ygg. Yellow is the \Lya\ emission line for \Bifrost. Green
is the CO(8--7) emission lines for \Ygg\ and \Loke. Blue and purple are
the center of the absorption feature 2.  The emitting components are
distributed within $\pm200$\,km/s of the systemic velocity, while the
absorbers are all blue shifted by $\sim$350\,km/s}.}
\label{fig:veldist}
\end{figure}

The MUSE spectrum of \Ygg\ (Fig.~\ref{fig:ions}) detects a rich variety
of ionised gas tracers. Overall, our spectrum is consistent with the Keck
spectrum of \citet{vernet01}, but now provides full spatial information
over the large field ($1'\times1'$) of view of MUSE.  Table~\ref{table:fluxes} lists
the velocity and integrated fluxes for the nine emission lines detected
at the position of \Ygg.  The flux is determined by integrating the
spectral line profile within $3\sigma$ of the noise. For our analysis, we adopt
the redshift of \HeII\ integrated over the peak of the continuum
emission as the systemic velocity, as it is the brightest non-resonant line in our spectra.  The \HeII\ emission line is
best fitted by a Lorentzian profile, with a centre corresponding to a
redshift of $2.92296\pm0.00001$ and a FWHM of $862\pm30$\,km/s.  The flux
within 3$\sigma$ is given in Table~\ref{table:fluxes}.  The other high
ionisation lines have profiles and redshifts similar to that of \HeII\
(Table~\ref{table:fluxes}) and Fig.~\ref{fig:veldist}.

\begin{figure*}[tbh]
\centering
\includegraphics[trim=0cm 0.5cm 0cm 0cm, clip=true, scale=0.65]{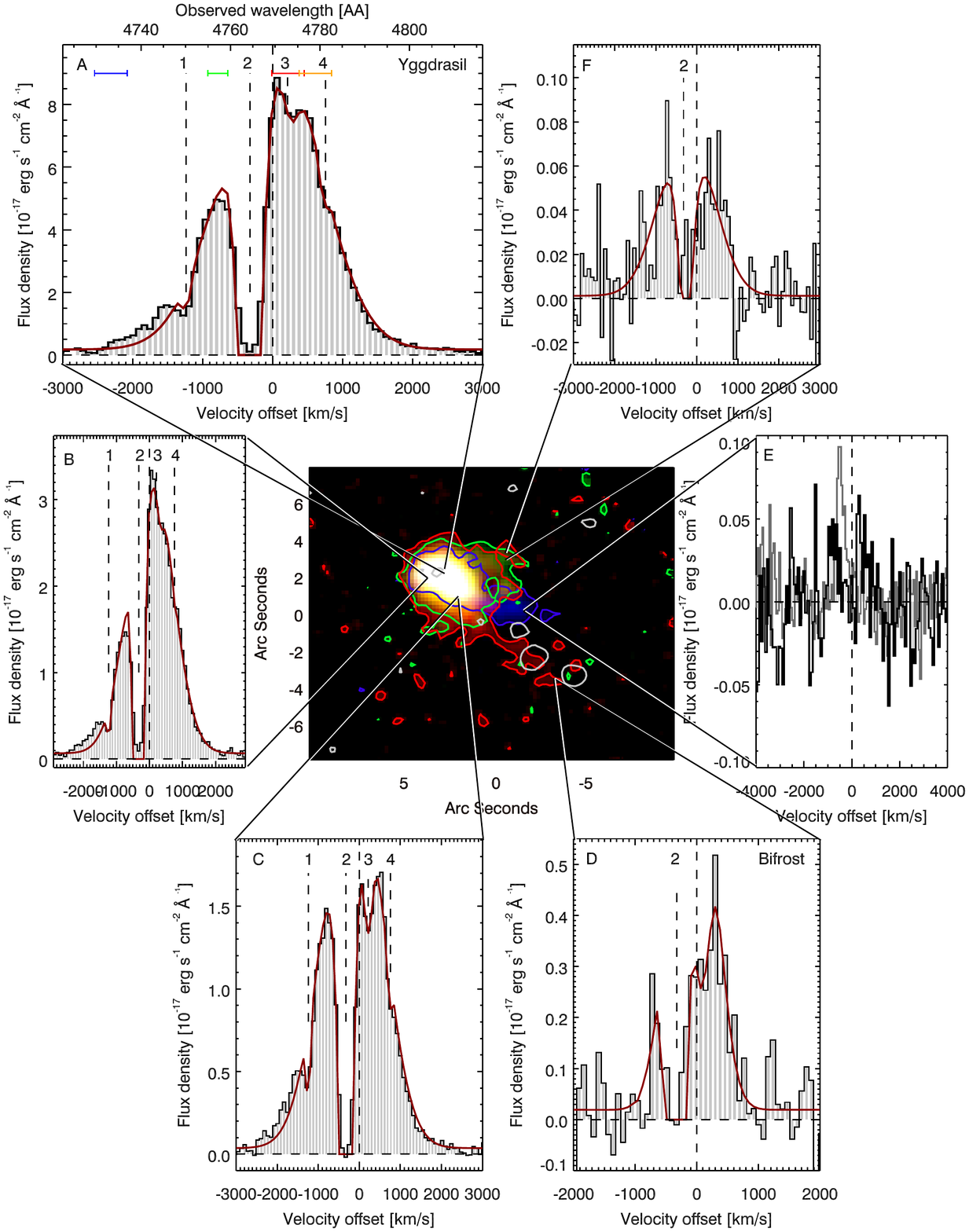}
\caption{{\small Overview of emission and absorption components at
different positions around MRC0943-242.  \textit{Middle Panel:} Composite
Red-Green-Blue image of \Lya\ and \HeII\ emission (see caption of
Fig.~\ref{fig:morph} for more details) with the ALMA \dustcont\ contours
overlaid in grey.  \textit{Panel A:} The \Lya\ line profile of the full
area of \Ygg\ (elliptical aperture of $1.3\arcsec\times0.7\arcsec$),
showing all four absorption components. Component 2 is the most prominent
and goes to zero intensity at its center. The spectrum has an
RMS of 0.3$\times$10$^{-17}$ erg s$^{-1}$ cm$^{-2}$ $\AA^{-1}$. The
blue, green, red and orange bars above the spectrum shows the range
in wavelength the channelmaps in Fig.~\ref{fig:channelmaps} have
been summed over. The blue correspond to panel A, green to panel B,
red to panel C and orange to panel C in Fig.~\ref{fig:channelmaps}.  \textit{Panel B and C:} The
\Lya\ profiles of two areas near the nucleus (circular $0.5\arcsec$
apertures) and have RMS of 0.1$\times$10$^{-17}$ erg s$^{-1}$
cm$^{-2}$ $\AA^{-1}$. These profiles likewise show signs of all four
absorption components.  \textit{Panel D:} The \Lya\ profile of \Bifrost\
(circular $1.2\arcsec$aperture and RMS of 0.1$\times$10$^{-17}$ erg s$^{-1}$) showing sign
of absorption component 2.  \textit{Panel E:} Spectrum extracted at the
\HeII\ emitting tail ($0.7\arcsec$ aperture), showing no sign of \Lya\
emission and an RMS of 0.02$\times$10$^{-17}$ erg s$^{-1}$
cm$^{-2}$ $\AA^{-1}$. The \HeII\ lines is over-plotted in grey.
\textit{Panel F:} The \Lya\ profile of the western extended \Lya\ emitting
gas ($0.4\arcsec$ aperture).  All \Lya\ velocity profiles have
been fitted with Voigt profiles superimposed on a Gaussian profile. These
fits are shown as the red curves over plotted on the spectra. }}
\label{fig:RGB}
\end{figure*}

The \Lya\ ($\lambda1215.7$\,\AA) line is by far the brightest and
most extended emission line in the MUSE data cube. It peaks at the
position of the AGN, but extends out to 80\,kpc (see bottom panel of
Fig.~\ref{fig:RGB}). Such extended \Lya\ haloes have been detected in
several other \HzRG s \citep[e.g.][]{vanojik96,reuland03,swinbank15}.
The \Lya\ halo in MRC0943-242 shows a linear filament, \Bifrost,
connecting \Ygg\ with the two brightest dust continuum components (\Thor\
and \Odin), strongly suggesting that they are physically connected
(see Fig~\ref{fig:RGB}).  The peak intensity of \Bifrost\ is $\sim18$
times fainter than that for the \Lya \ detection of \Ygg\ (see
panel A and D of Fig.~\ref{fig:RGB}).

The \Lya\ emission on the nucleus exhibits four absorption features
in our data.  We denote these absorption features as 1, 2, 3 and 4
where the bluest feature is 1, middle are 2 and 3, and the reddest
absorption feature 4 \citep[in accordance with the labelling
used by][see panel A of Fig.~\ref{fig:RGB}]{jarvis03}.
Furthermore, component 2 is related to the strong \CIV\ absorption
observed against the nucleus \citep{binette00}.  By centering the
\Lya\ velocity profile at the fitted central velocity of \HeII\,
we find that the bottom of the absorption trough is at a redshift off
$z_{\text{abs}}=2.91864\pm0.00002$. By fitting a Voigt-profile to the
absorption trough, we estimate a  column density of the neutral gas for
component 2 of \NHI \ $=1.3\cdot10^{19}$\,cm$^{-2}$ and a Doppler width
of 64\,km/s (see Table~\ref{table:voigt}) which are comparable to the
values determined by \citealt{binette00}.  We note that because the line
is saturated, this estimate is very uncertain and likely a lower limit
to the true value.  For components 1, 3 and 4, we find good agreement
with the values in \citet{jarvis03} for the column densities and velocities.

We observe two absorption troughs in the \CIV\ emission line; Voigt
profile fitting to the two features reveals that they are caused by gas
on the line of sight at redshift $z_{\text{abs}}=2.91940\pm0.00007$
and $2.91945\pm0.000061$ and column densities of \NCIV \
$=3.4\cdot10^{14}$\,cm$^{-2}$ and $=6.9\cdot10^{14}$\,cm$^{-2}$, higher
than $3.2\cdot10^{14}$\,cm$^{-2}$ found by \citealt{binette00}. 

\begin{table}      
\centering          
\begin{tabular}{l c c c }
\hline\hline  
Line &  \NHI \    & b      & $z_{\text{abs}}$\\
        & \cmtwo & km/s &                           \\
\hline
\textbf{\Ygg} & & \\
\Lya\ absorber 1         & $5.0\cdot10^{13}$ & 84 & $2.90689\pm0.00032$ \\
\Lya\ absorber 2         &$1.3\cdot10^{19}$ & 64 & $2.91864\pm0.00002$ \\
\Lya\ absorber 3         &$3.7\cdot10^{13}$ & 140 & $2.92641\pm0.00023$ \\
\Lya\ absorber 4         &$1.7\cdot10^{13}$ & 33 & $2.93254\pm0.00007$ \\
\CIV\ absorber 2         &$3.4\cdot10^{14}$ & 94 &  $2.91940\pm0.00007$ \\
                                   &$6.9\cdot10^{14}$ & 100 & $2.91945\pm0.00006$ \\

\hline 
\textbf{\Bifrost} & & \\
\Lya\ absorber 2  & $7\cdot10^{18}$ & --- & $2.91841\pm0.00045$\\ 
\hline 
\end{tabular}
\caption{{\small Parameters determined by fitting a Voigt profiles to
the absorption troughs. Fits are performed on the absorption troughs
in the \Lya \ line observed for both \Ygg \ and in-between \Odin \ and
\Thor. The \CIV \ line has been fitted with two Voigt profiles with the
same fitting parameters.}}
\label{table:voigt}
\end{table}

\begin{figure*}[tbh]
\centering
\includegraphics[trim=0.6cm 9.6cm 6.3cm 0.8cm, clip=true, angle=90, scale=0.8]{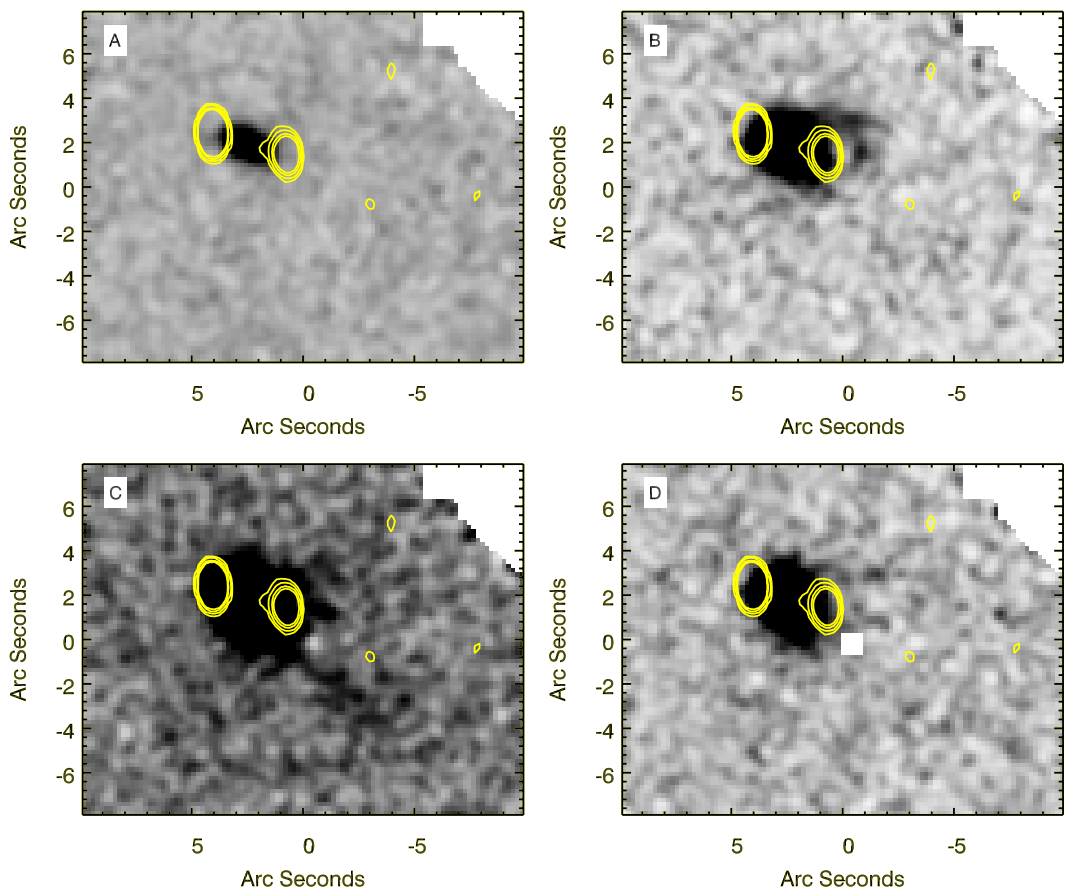}
\caption{{\small \Lya\ Moment-0 maps summed over the wavelength ranges illustrated with the coloured bars in panel A of Fig.~\ref{fig:RGB}: 
	\textit{Panel A:} sum over the blue bar ($\lambda_{\text{obs}}=4732.5-4747.5$).
	\textit{Panel B:} sum over the green bar ($\lambda_{\text{obs}}=4754.3-4758.8$).
	\textit{Panel C:} sum over the red bar ($\lambda_{\text{obs}}=4768.8-4776.2$). 
	\textit{Panel D:} sum over the orange bar ($\lambda_{\text{obs}}=4778.8-4793.8$)}}
\label{fig:channelmaps}
\end{figure*}

\begin{figure*}[tbh]
\centering
\includegraphics[trim=0.6cm 9.6cm 6.3cm 0.8cm, clip=true, angle=90, scale=0.8]{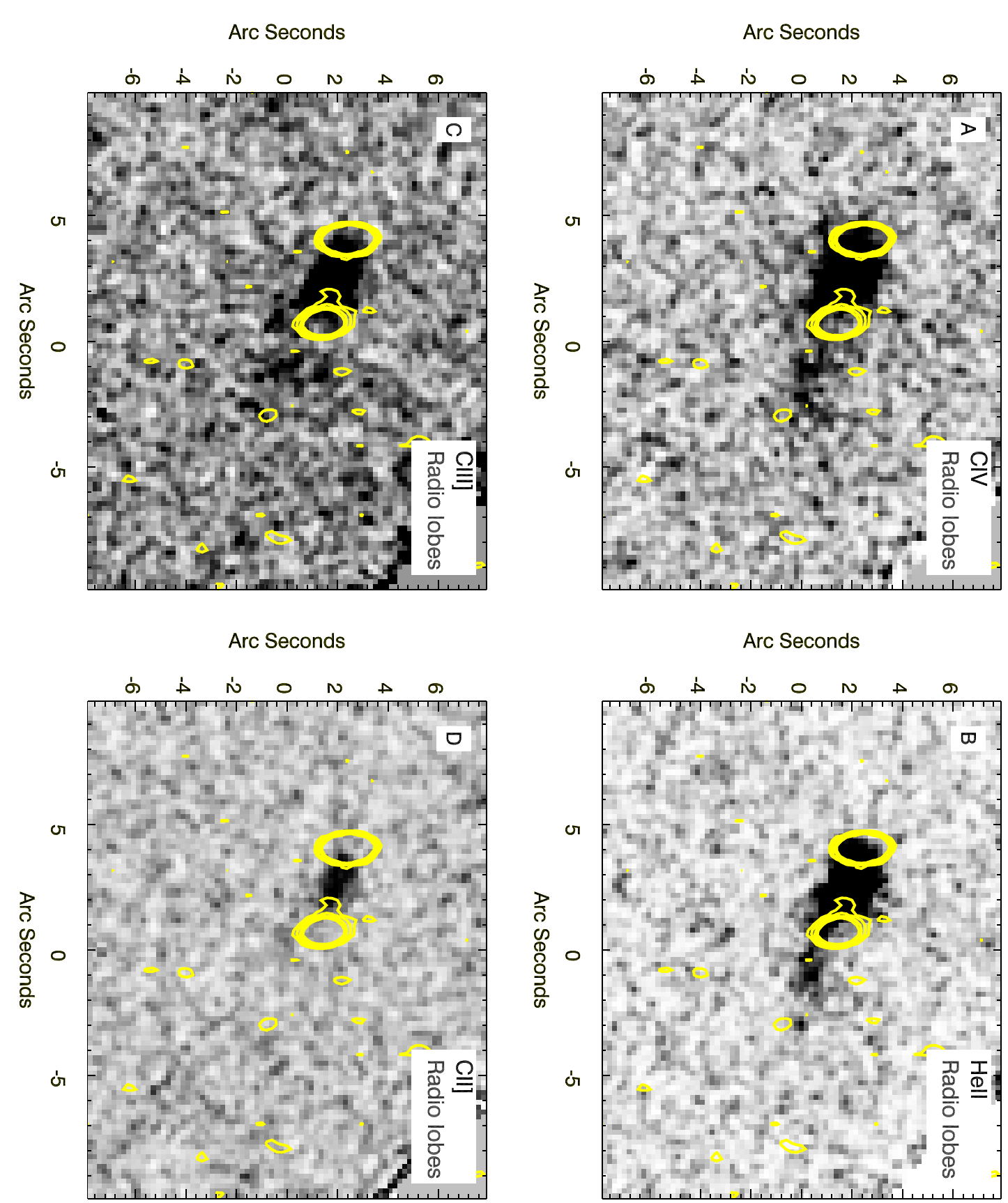}
\caption{{\small Continuum subtracted moment-0 maps of the \CIV, \HeII, \CIII\ and \CII\ emission lines. The continuum is determined from the nearby line free channels for each line. 
	\textit{Panel A:} sum over the \CIV\ emission line for the wavelength range $\lambda_{\text{obs}}=6050-6090$\,\AA.
	\textit{Panel B:} sum over the \HeII\ emission line for the wavelength range $\lambda_{\text{obs}}=6422-6430$\,\AA \ and is the same image as the blue colour of the middle panel of Fig.~\ref{fig:RGB}.
	\textit{Panel C:} sum over the \CIII\ emission line for the wavelength range $\lambda_{\text{obs}}=7455-7485$\,\AA.
	\textit{Panel D:} sum over the \CII\ emission line for the wavelength range $\lambda_{\text{obs}}=9071-9171$\,\AA. 
	Over-plotted with yellow contours in all panels is the VLA radio map, as a reference.}}
\label{fig:carbon_he}
\end{figure*}

Stepping through the cube towards increasing wavelengths reveals
the complexity of the emission in the MRC0943-242 system (see
Fig.~\ref{fig:RGB}, \ref{fig:channelmaps} and \ref{fig:carbon_he}):

\noindent
\textbf{$\bullet$ $\lambda_{\text{obs}}\sim4732-4747$\,\AA:}
Considering the \Lya\ emission line profile at \Ygg\ (see panel A of
Fig.~\ref{fig:RGB}); the emission furthest to the blue side of the line
profile is spatially centred around the radio source and approximately
within the region containing the radio emission (see also panel
A Fig.~\ref{fig:channelmaps}).

\noindent
\textbf{$\bullet$ $\lambda_{\text{obs}}\sim4747-4754$\,\AA:} As we move
towards the red, over the wavelength range of absorption feature 1,
we see strong nuclear emission.

\noindent
\textbf{$\bullet$ $\lambda_{\text{obs}}\sim4754-4759$\,\AA:} Moving
further to the red between absorption features 1 and 2, we observe
a closed region of bubble-like emission reaching $\sim20$\,kpc
out, surrounding the western radio lobe (see panel B of
Fig.~\ref{fig:channelmaps}).  The absorption components 1 and 2 are
visible against this emission across the region of \Ygg\ (see panel A-C
of Fig.~\ref{fig:RGB}).

\noindent
\textbf{$\bullet$ $\lambda_{\text{obs}}\sim4759-4769$\,\AA:} As we step
further to the red, through the wavelength range of absorption feature
2, the \Lya\ emission becomes fainter -- down to our surface brightness
detection limits.  We note that this is in agreement with the \Lya\
profiles of all the extended emission -- when we have sufficient surface
brightness to have detectable line emission, we see absorption component
2 which has zero intensity in the line core.  This includes the \Lya\
emission from \Bifrost\ (see panel D in Fig.~\ref{fig:RGB}) and the
\CIV\ emission from \Ygg. The absorption component 2 in \Bifrost\ is at a 
redshift of $z_{\text{abs}}=2.91841\pm0.00045$, consistent with that in \Ygg\ (see panel D of Fig~\ref{fig:RGB}).
This suggests that the absorption feature in the spectrum of 
\Bifrost\ is due to absorption component 2, meaning that  
absorber 2 has an extent of at least 65\,kpc from \Ygg.
The data suggest a lower column density of absorber 2 in \Bifrost\ of
\NHI\,$\sim7\cdot10^{18}$\,cm$^{-2}$. We warn that this \NHI\ is particularly uncertain
due to the low S/N of the data, and the degeneracy between \NHI\ and the b-parameter. 
Deeper and higher resolution data are required to trace column density variations across the system.
 We also note that in this wavelength range, we
observe high ionisation emission lines of \HeII, \CIV, and \CIII\ from
gas extending over 7\arcsec\ or $\sim60$\,kpc to the west (see panel
F, the green contours of the central panel of in Fig.~\ref{fig:RGB}
and the \CIV\ and \HeII\ moment-0 maps in panel A and B of
Fig.~\ref{fig:carbon_he}).  Fainter, less extended emission is also
observed to the east of the nucleus.  Deeper data will be needed to
determine the exact distribution of absorption component 2; however,
since the line core is black throughout the extend of the cube, we
can conclude that it is extended with a unity covering factor.

\noindent
\textbf{$\bullet$ $\lambda_{\text{obs}}\sim4769-4776$\,\AA:} At
wavelengths to the red of absorption feature 2, and containing the
absorption feature 3 (see red contours of Fig.~\ref{fig:RGB} and
panel C in Fig.~\ref{fig:channelmaps}), we observe the emission from
\Bifrost. Where the \Lya\ emission is sufficiently bright, we see 
absorption component 3 superposed on the line emission (see panel
A-C of Fig.~\ref{fig:RGB}). 

\noindent
\textbf{$\bullet$ $\lambda_{\text{obs}}\sim4779-4794$\,\AA:} Further
to the red - at longer wavelengths than the range of absorption
feature 3, component 4 shows up as a small trough (see panel A-C of
Fig.~\ref{fig:RGB} and panel D of Fig.~\ref{fig:channelmaps})
before we again start to see circum-nuclear emission all contained within
the radius encompassing the radio lobes.

\noindent
\textbf{$\bullet$ $\lambda_{\text{obs}}\sim6025-6125\ {\rm and}\
6422-6430$\,\AA:} Moving to wavelengths of the \CIV\ and \HeII\ emission,
we do not have sufficient S/N to split these lines into velocity channels
like for \Lya. Instead, we present line maps of the central parts of these
lines in panels A and B of Fig.~\ref{fig:carbon_he}. Both lines display
a similar morphology, which is also seen (albeit at lower S/N) in \CIII\
(panel C of Fig.~\ref{fig:carbon_he}).  We detect a tail of \CIV\ and
\HeII\ emission to the WSW (panels A and B of Fig.~\ref{fig:carbon_he});
however this tail is not spatially aligned with the \Lya\ bridge
(\Bifrost) connecting \Ygg\ with \Freja, \Thor\ and \Odin. Interestingly,
the \CIV\ and \HeII\ tail is not seen in \Lya\ (see the blue contours
of the middle panel and panel E of Fig.~\ref{fig:RGB}).  Fitting a
Lorentz profile to this \HeII\ emission line as well yields a redshift
of $z=2.916$ which is consistent with the emission being at the same
velocity offset as absorption component 2. This suggest that this \CIV\
and \HeII\ emitting gas is also seen in \Lya, but that this is absorbed
completely by component 2.
\\\\
The large wavelength coverage of MUSE allows us to detect other fine
structure lines (see panel B-I of Fig.~\ref{fig:ions}).
The three doublets \NV, \OIII\ and \CIII\ are well detected at a
$\geq6\sigma$ level.  The \NV \ doublet is spatially resolved and
reveals two spectral lines of similar intensity. The \OIII\ doublet is
blended, but there is some indications of the doublet structure with
the $\lambda1660.8$\,\AA\ line being a shoulder in the blue side of
the \OIII~$\lambda1666.1$\,\AA\ line. We also detect a broad \SiIV \
component, a weak \NIV \ line at a $2.3\sigma$ level, and a \CII\ line
at a $5.5\sigma$ level at the edge of the band.  The \CII\ emission line
is a blend of multiple lines, which explains why the line peaks on the
red side of $\lambda2326$\,\AA.  Table~\ref{table:fluxes} lists the line
parameters for these fine structure lines over the circum-nuclear region
(see Fig.~\ref{fig:ions}).

All the emission from \NV, \SiIV, \NIV, \OIII, and \CII\ lies within
the region emcompassed by the radio lobes. The \CII\ emission is a
particularly spectacular example as it delineates both the extension
and direction of the radio jet and lobes (see panel D of
Fig.~\ref{fig:carbon_he}).

\subsection{Stellar mass}\label{sec:stellarmass}

The wavelengths covered by \textit{Spitzer} probe the rest-frame near-
and mid-IR part of the spectrum, which is believed to be the best
tracer of the stellar-mass with almost no contributions from the thermal
emission from the AGN \citep{seymour07}. 
\Ygg\ has a well sampled SED with detections at
$3.6\,\mu$m, $4.5\,\mu$m, $8.0\,\mu$m $16\,\mu$m, $24\,\mu$m, and upper
limits for $5.8\,\mu$m, $70\,\mu$m and $160\,\mu$m. \citealt{seymour07}
and \citealt{debreuck10} fit a toy model to these points, composed of
four components: an elliptical galaxy composite stellar population (CSP)
and three blackbody (BB) components with temperatures of 60\,K, 250\,K
and between 500-1500\,K (for more details see \citealt{seymour07} and
\citealt{debreuck10}). This SED fitting results in an $H$-band stellar
luminosity of $L_{H}^{\text{stellar}}=2.5\cdot10^{11}$\,L$_{\odot}$,
which is converted to a stellar mass of $1.7\cdot10^{11}$\,M$_{\odot}$ by
assuming a dust-free, passively evolving elliptical galaxy which started
its formation at $z=10$ with no recent episodes of star-formation. The
SED allows for a good decomposition of the stellar and hot-dust
dominated components, so the derived stellar mass is expected to
be reliable \citep[see ][for a more detailed discussion on the
uncertainties]{seymour07}.

Examining the IRAC $3.6\,\mu$m and $4.5\,\mu$m image again, now with
the knowledge of the presence of \Freja, \Thor\ and \Odin, we find
stellar emission at the position of \Thor, with an extension towards
\Freja\ and \Odin\ (see Fig.~\ref{fig:morph} top left and
Table~\ref{table:photometry}). Also for \Thor\ are the stellar and
hot-dust emission well separated in the SED.  We scale the stellar
mass found for \Ygg, to the emission seen in the IRAC $4.5\,\mu$m map
(Table~\ref{table:masses}), yielding a stellar mass of $1.0\times
10^{11}$\,M$_{\odot}$. For \Loke, we accordingly derive an upper limit to the stellar mass of $\lesssim 6.2\times 10^{9}$\,M$_{\odot}$.

\begin{table}  
\centering          
\begin{tabular}{c c c c c }
\hline\hline 
Instrument  & wavelength & $S_{\text{\Ygg}}$ & $S_{\text{\Thor}}$ &$S_{\text{\Loke}}$\\ 
                   &  $\mu$m     &  $\mu$Jy              &   $\mu$Jy              &  $\mu$Jy        \\ 
\hline                    
IRAC 1 & 3.6 & $21.9$   & $9.02$ & <2.10 \\
IRAC 2 & 4.5 & $36.1$   & $21.0$ & <1.32 \\
IRAC 3 & 5.8 & ---          & ---         & --- \\
IRAC 4 & 8.0 & 30.2 & $ <21.4$ & --- \\
MIPS    & 24  & $468\pm40$ & $180\pm40$ & ---\\
\hline
\end{tabular}
\caption{{\small The photometric IRAC and MIPS points for \Ygg, \Thor\ and \Loke\ after de-blending. \Thor\ is not detected in the IRAC 3 image,
however the flux at the position is influenced by an image artefact
from a nearby star in the field. Extracting even an upper limit at this
position is therefore not possible. \Loke\ is not 
detected in any of the IRAC images; we quote the 3$\sigma$ upper limits \citep{wylezalek13}.}}
\label{table:photometry}
\end{table}

\begin{table}      
\centering          
\begin{tabular}{l c c c}
\hline\hline       
Component & \multicolumn{2}{c}{Position} & $S_{235\,\text{GHz}}$ \\ 
                   &             RA & dec                  &     mJy                          \\ 
\hline                    
Yggdrasil & 09:45:32.769 & -24.28.49.29 & $0.8\pm0.2$ \\ 
Odin        & 09:45:32.222 & -24.28.55.06 &  $2.0\pm1.0$ \\ 
Thor        & 09:45:32.386 & -24.28.54.05 & $1.9\pm0.8$  \\ 
Freja       & 09:45:32.445 & -24.28.52.55 & $0.6\pm0.3$  \\ 
Loke       & 09:45:32.072 & -24.28.56.94 &  $<0.3$ \\          
\hline                  
\end{tabular}
\caption{{\small Positions and \dustcont\ fluxes for all components of MRC0943-242. As no dust continuum is observed for the position of \Loke\ we infer a $3\sigma$ upper limit of the \dustcont\ of three times the rms. The uncertainty includes the 15\% flux calibration error.}}
\label{table:data}
\end{table}

\subsection{Disentangling the SED}
\begin{table}
\label{tab:results_SSED}
\begin{tabular}{lcc}
\hline
Component & L$_{\text{IR}}$                & perc. \\
                    & $[10^{12}$\,L$_{\odot}]$ &          \\
\hline \hline
Total (\Ygg+\Freja+\Thor+\Odin) &17.2 & 100 \\
\Freja+\Thor+\Odin\ (SB) & 8.4 & 49 \\
\Ygg\ (AGN) & 7.5 & 43 \\
\Ygg\ (SB) & 1.3 & 8 \\
\hline
\end{tabular}
\caption{{\small Infrared luminosities of each component, determined from our SED presented in Fig~\ref{fig:SED}.}}
\label{table:SEDres}
\end{table}

\begin{figure}[h]
\includegraphics[trim=0.5cm 0.4cm 0cm 0cm, clip=true, scale=0.54]{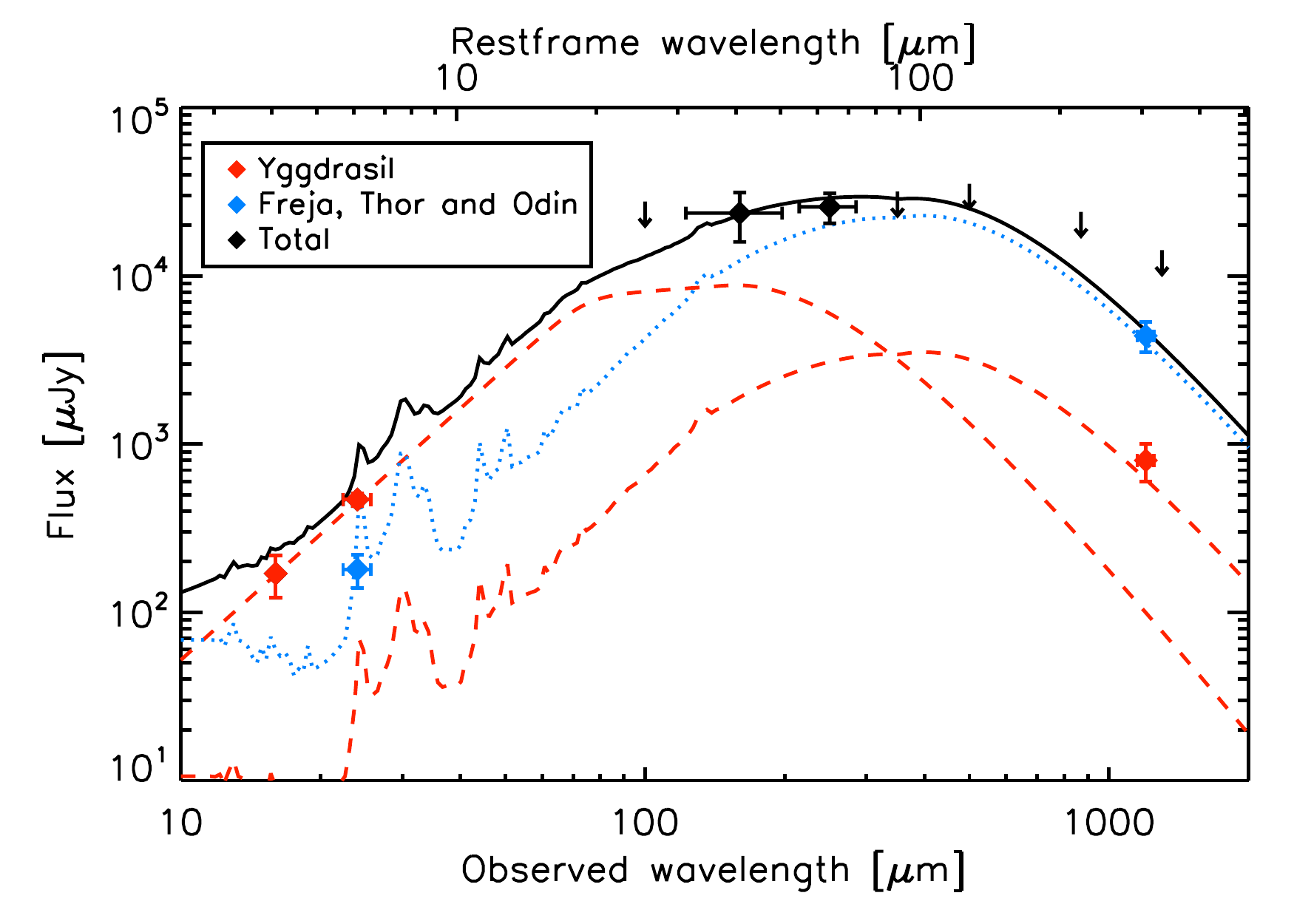}
\caption{{\small The spectral energy distribution of MRC0943-242. The
spatial resolution of the ALMA data allows us to disentangle the SED in
the AGN heated component (red curve) and star formation heated component
(blue curve). The sum of these two components is illustrated by the
black curve.}}
\label{fig:SED}
\end{figure}

\Ygg\ has a well-sampled SED with IRAC, MIPS, SPIRE ($250$\,$\mu$m,
$350$\,$\mu$m and $500$\,$\mu$m) and PACS ($24$\,$\mu$m, $100$\,$\mu$m
and $160$\,$\mu$m) from previous studies \citep{debreuck10, drouart14}.
The PACS 160\,$\mu$m image \citep[see Appendix C of ][]{drouart14} reveals
an elongated source, spanning from the position of \Ygg \ towards \Freja,
\Thor\ and \Odin.  Likewise, when returning to the IRAC 3.6\,$\mu$m
and 4.5\,$\mu$m images (see panel A and B of Fig.~\ref{fig:morph}),
multiple components show up corresponding to \Ygg \ and \Thor\ (see
\S\ref{sec:stellarmass}). While these sources can be separated in
the \textit{Spitzer} data, they are blended in the \textit{Herschel}
data. However, the high spatial resolution and sensitivity of the ALMA
data allows us to disentangle the AGN and star formation heated components
in the SED at 235\,GHz. We conclude that the star-formation,
as probed by the dust emission, is taking place at a distance of
48-65\,kpc from the AGN in three smaller aligned components (\Freja,
\Thor\ and \Odin). Table~\ref{table:photometry} lists the de-blended
MIPS and IRAC photometry points.

Relying on the method of \citealt{drouart14} by fitting an SED with
AGN and starburst components, we now go further, and use the sub-arcsec
resolution provided by the new dust continuum image to further disentangle
the relative contribution of AGN and star formation {\it spatially}.
Figure~\ref{fig:SED} shows the composite SED of the system, the
three components being two spatially distinct positions with \textit{i}) \Ygg\ \textit{ii}) and
\Freja, \Thor\ and \Odin\ photometry as red and blue points respectively.
The unresolved far-IR {\it Herschel} detections and upper limits are
shown as black points.  It is important to note that each component
is fitted individually, and that the black points are not used in the
fit. However, the total SED and unresolved data (in black) provide
a strong constraint on the total integrated flux from all components;
all solutions where the sum is inconsistent with the {\it Herschel}
photometry are excluded. In practice, the starburst component in
\Freja, \Thor\ and \Odin\ is well constrained (blue dotted line in
Fig.~\ref{fig:SED}). However, the AGN component cannot simultaneously
fit both the {\it Spitzer} and ALMA photometry without exceeding the
{\it Herschel} photometry. We are therefore forced to add an additional
starburst component to fit the \Ygg\ ALMA point.  However, with only a
single point, the starburst luminosity is only loosely constrained.

We clearly see from the SED fitting that \Thor\ is star-formation
dominated, with the MIPS 24\,$\mu$m and 235\,GHz reproduced by the
starburst model from DecompIR \citep{mullaney11}.  \Ygg\ represents
a composite of emission from an AGN and star formation as reported
earlier \citep{drouart14}. However, the relative contribution to
the IR luminosity from the AGN and the star formation is different;
favoring the AGN as the main contributor. In fact, the AGN emits
$\sim$40\% of the IR luminosity of the entire system ($L_{\rm total}^{\rm
IR}=1.7\times10^{13}$\,L$_{\sun}$), the remaining $\sim 60$\% is star-formation
split between \textit{i}) \Ygg\ and \textit{ii}) \Freja, \Thor\ and \Odin\ at $\sim$10\% and
$\sim$50\% levels (Table~\ref{tab:results_SSED}).  According to the
\cite{Kennicutt98} relation, \Ygg\ and \Freja, \Thor\ and \Odin\ have
SFR of $\sim 200$ and $\sim 1400$\,M$_{\odot}$\,yr$^{-1}$, respectively. Which,
in turn, corresponds to a sSFR of $\sim$10$^{-9}$ and 10$^{-8}$\,yr$^{-1}$.

\section{Discussion}\label{sec:disc}

The wealth of information provided by the MUSE and ALMA data, even
with rather modest integration times for both sets of data, paints a
complicated picture from the warm ionised gas to the cold molecular gas
and dust.  We now discuss how the mass is distributed in each component
and what the relationship between the components is.

\subsection{The nature of the gas and dust} 

The MRC0943-242 system has a complex distribution of the ionised and
neutral gas. The high ionisation lines are distributed mainly around
the nucleus and are confined within the radio lobes.  The exception
to this is the extended \HeII, \CIV, and \CIII\ emission which is most
extended to the west -- beyond the western radio lobe (see panel
A-C of Fig.~\ref{fig:carbon_he}). The fact that metals are detected
within this part of the halo suggests the gas is not pristine, while
the high ionisation state of the gas is consistent with photoionisation
by the AGN, indicating that there is a cone of ionisation along
the direction of the radio jet but extending well beyond the radio
lobes. The lack of very extended \Lya\ emission in the outer region of
the ionisation cone is most likely caused by \HI\ gas distributed over
large scales absorbing the emission (absorption component 2).  The inner
regions of the ionisation cone, i.e. the region that lies within or
just beyond the radio lobes, shows significant \Lya\ emission (and also
absorption from component 2).  Resonance scattering shifts the frequency
of the \Lya\ photons and allow them to 'leak' out on both sides of the
absorption feature and broaden the line.  In fact, throughout the region
of \Ygg, we see all absorption components either in all places (as for
components 1, 2 and 3) or in particular regions (as for component 4).
Since the absorption features are only rarely reaching zero intensity
in their cores, the neutral gas responsible for absorption systems 1,
3, and 4 are mixed with \Lya\ line emitting regions (although some of
it overlies the \Lya\ emission).  Absorption component 2 must cover
the full \Lya\ emitting region, including that out to $\sim$60\,kpc,
which is the projected distance to \Bifrost.  Since this absorption
contains no flux in the line core (it is dark), the absorber must have
unity covering factor and be optically thick.  This being the case,
our estimate of the \HI\ column density is a lower limit.

This situation is somewhat akin to the large extended \Lya\
absorber on the line of sight to the \HzRG\ TN J1338-1942 at $z=4.1$
\citep{swinbank15}. In this case the neutral gas reservoir extends at
least 150\,kpc away from the core of the \HzRG. To detect this extended
emission, \citet{swinbank15} obtained a much deeper integration and hence
significantly more extended \Lya\ emission and absorption. However,
unlike the deepest absorption feature in TN J1338-1942, the offset
velocity of the absorption component 2 in MRC0943-242, is rather more
modest, only a few 100\,km\,s$^{-1}$ compared to 1200\,km\,s$^{-1}$.
Deeper data for MRC0934-242 may reveal a greater extent and
more complex structure of the emitting and absorbing gas.

In addition to the ionisation cone and the general distribution of the
\Lya\ emission around the central regions of \Ygg\ ($M_*\simeq 1.7\times10^{11}$\,M$_{\odot}$), there is a bridge of
material connecting the IRAC continuum emission of \Thor\ and beyond ($M_* \simeq 1.0\times10^{11}$\,M$_{\odot}$).
\cite{ivison08} also found multiple components for the \HzRG\ 4C60.07 at
$z=3.8$, likewise connected by a bridge of gas. They interpret this bridge
as a plume of cold dust and gas in a tidal stream between two interacting
galaxies. The interaction between the two galaxies is believed to be
the trigger of the starburst and through this exhaust the molecular gas
supply in the AGN host galaxy.  One scenario which could have triggered
the high star formation in \Freja, \Thor\ and \Odin\ and perhaps the
AGN activity in \Ygg\ is, like for 4C60.07, interaction between two
galaxies. In this scenario, the two galaxies interact tidally where gas
flows between the galaxies trigger the AGN activity in \Ygg, and creates
tidal tails of gas which may emit \Lya. The tidal interactions may then
ignite the string of components \Freja, \Thor\ and \Odin\ resulting in a
SFR $\sim1400$\,M$_{\odot}$/yr.  However, the CO(8--7) emission
line from \Ygg\ is very narrow and dynamically quiescent. It shows no
signs of dynamical motion, implying that this gas is not related to the
tidal interaction. The CO gas appears to be quite highly excited, as may be expected for gas near the central AGN. However, we remark that the ATCA upper limit on the CO(1-0) emission is rather shallow, so we cannot put strong constraints on the CO gas excitation.

Along the same projected line as the bridge and dust emitting regions,
we find CO(8--7) emission in \Loke\ which is not detected in any optical
or dust continuum counterpart. The tentative detection of CO(1--0)
at the position of \Loke\ is consistent with the presence
of a large reservoir of molecular gas. However, it is not clear from
the observations at hand how extended this reservoir is, if it reaches
across \Freja, \Thor, \Odin\ and \Loke, tracing one compact region, or
if the reservoir is only at the location of \Loke. The tentative CO(1-0) 
detection suggests a lower excitation level than \Ygg, but more sensitive 
low-$J$ observations are needed to confirm this. The only thing
known is that this region contains dynamically quiet but likely highly excited
gas component $\sim90$\,kpc away from the nearest stellar emission and with
a relatively low velocity compared to the systemic velocity of the AGN.
A clue to its nature lies in the alignment with the AGN and starburst
components along the bridge.  If this represents an accretion flow of
galaxies and material along this direction, the gas would experience
tidal forces and accretion shocks when the galaxies get close (or enter
the halo of \Ygg). The absorption component 2 reveals the presence of a
large amount of neutral gas at a radius $>60$\,kpc from the AGN, however
the impact on this gas from merging galaxies or accreting gas is unknown.
It may be that as gas enters the halo, it will encounter the large scale
neutral gas, perhaps shocking it as it penetrates. Only more sensitive
studies of the molecular gas in \Loke\ may this impact be revealed.

\subsection{Distribution of masses}

To understand the nature of the MRC0943-242 system, it is
important to know how the mass is distributed in the system.
Following \citet{debreuck03}, we estimate the ionised gas
masses of \Ygg. The mass of the ionised gas is given by
$M_{\text{ion}}=10^9(f_{-5}L_{44}V_{70})^{1/2}$\,M$_{\odot}$,
where $f_{-5}$ is the filling factor in units of $10^{-5}$,
$L_{44}$ is the \Lya\ luminosity in units of $10^{44}$\,erg/s and
$V_{70}$ is the volume in units of $10^{70}$\,cm$^3$. We assume
a filling factor of $10^{-5}$ \citep{mccarthy90} and a volume of
$5\arcsec \times 2.5\arcsec\times 2.5 \arcsec$, and therewith find
M$_{\text{ion}}=5.2\times10^{8}$\,M$_{\odot}$.  This is a lower limit to
the total mass since the absorption in the \Lya\ emission is significant.
Since the distribution of the absorption components 1, 3, and 4 are
only delineated by the \Lya\ emission, it is unknown how this gas is
distributed, meaning we are most likely underestimating their column
densities as the components do not fully cover the emission.  This means
we cannot estimate their contributions to the mass budget.

Absorption component 2 is observed in all the \Lya\ line profiles
across the \Lya\ emitting regions and must therefore cover the entire
\Lya\ emitting region out to at least $\sim60$\,kpc.  Assuming it
is distributed as a spherical shell, we estimate the total mass
of absorption component 2 again following \citet{debreuck03},
$M_{\text{neutral}} \ga 3.8 \times10^{9}(R/60\,\text{kpc})^2
(N_{\text{HI}}/10^{19})$\,cm$^{-2}$\,M$_{\odot}$, where $R$ is the radius
of the neutral gas and \NHI\ is the \HI\ column density.  We note that
because the absorption is heavily damped \citep{jarvis03}, our column
density estimate is likely a lower limit of the true column, and our
mass estimate therefore a lower limit.

\begin{table}  
\centering          
\begin{tabular}{l c c }
\hline\hline       
Mass  & \Ygg & \Freja/\Thor/\Odin/\Loke \\ 
           &  M$_{\odot}$        &  M$_{\odot}$           \\ 
\hline                    
M$_{\text{stellar}}$ & $1.7\times10^{11}$ & $1\times10^{11}$\\
M$_{\text{ion}}$ & $5.2\times10^{8}$ & --- \\
M$_{\text{neutral}}$ & $3.8\times10^9$ & --- \\
M$_{\text{mol}}$ & $6\times10^{10}$ & $2.3\times10^{10}$\\
\hline
\end{tabular}
\caption{{\small The stellar mass, ionised-, neutral- and molecular gas mass for \Ygg\ and \Thor.}}
\label{table:masses}
\end{table}

Using the tentative CO(1--0) detection at the position of \Loke, and
choosing a conservative $\alpha_{\text{CO}}=0.8$\,M$_{\odot}$\,K\,km\,s$^{-1}$\,pc$^{-2}$
\citep{downes98} we estimate a molecular gas mass of
$2.3\times10^{10}$\,M$_{\odot}$ for \Loke.  \cite{emonts11}
discovered a tentative 3$\sigma$ CO(1--0) detection $\sim60$\,kpc
North-East of \Ygg, and estimate a molecular gas mass of
$M_{\text{H$_2$}}=6\times10^{10}$\,M$_{\odot}$, using the same value
for $\alpha_{\text{CO}}$.

It is clear from comparing the four mass estimates
(Table~\ref{table:masses}) that the most of the mass is in the form of
stars and molecular gas, although the contribution of the neutral shell
of gas may also be significant.

\subsection{Ionisation mechanism}

The high ionisation lines such as \HeII, \CIV, and \CIII\  are extended
beyond the western lobe and asymmetric on large scales (see
panel A-C of Fig.~\ref{fig:carbon_he}). The simplest interpretation is a
strong asymmetry in the gas distribution.  This hypothesis is supported
by the hard photons, which are most likely responsible for the
ionisation of the gas, causing the asymmetry in the emission from the
circum-nuclear gas. Moreover, the ALMA data show a similar asymmetry in
the distribution of both the dust and CO emission. However, this emitting
gas lies outside of the ionisation cone and can therefore not be affected.

The detected emission lines from \Lya, \CIV, \HeII, \CIII \
and \CII\ can be used as probes for the ionisation mechanisms. Both
the AGN and the far-UV field of vigorous star-formation can ionise
the gas. The line ratios of \Lya, \CIV\ and \HeII\ can be used to
determine the contribution from stellar photoionisation (Fig.~4 of
\citet{villar-martin07}). We find that the observed (absorption corrected)
line ratios of \Ygg\ (Table~\ref{table:fluxes}) are consistent with
$\omega=0$, i.e. pure AGN photo-ionisation. Similarly, we can use
the \CIV/\HeII, \CIII/\HeII, \CIV/\CIII, \CII/\CIII, \CIV/\Lya \ and
\CIII/\Lya \ flux ratios along with the diagrams of \citealt{debreuck00}
(their Fig.~13, adapted in Fig~\ref{fig:lineratios}). This
comparison shows that the gas within the radio lobes of \Ygg, is likely 
ionised by a combination of photo-ionisation ($\sim$70\%) and shock
+ precursors ($\sim$30\%). \CII\ emission is an especially sensitive
tracer of shock-ionised gas \citep{best00, debreuck00}. Our MUSE data
allow us to produce a \CII \ narrow-band (moment-0) image (see panel D
of Fig.~\ref{fig:carbon_he}). Interestingly, this image indeed reveals
that the \CII \ emission is confined within the radio lobes, as would be
expected if this line is dominated by shock ionised gas. Deeper MUSE data
would be needed to check if this is not just due to limited S/N. This
also limits us to derive line ratios in the extended halo (near \Bifrost),
where we detect \Lya, but no other emission lines. Deeper data would
allow to check if the gas at $\sim 80$\,kpc is still ionised by the AGN,
or has a more important contribution from stellar photo-ionisation or shocks.

\begin{table*}      
\centering          
\begin{tabular}{l l c c c c c }
\hline\hline       
\HzRG  & $z$ & CO detection & Separation$^a$ & CO/dust association  & Reference \\ 
\hline                    
MRC0114-211  & 1.402 & CO(1--0) & 35\,kpc & no & \cite{emonts14}; Nilsson et al. in prep.\\
MRC0152-209  & 1.921 & CO(1--0)/CO(6--5) & 10\,kpc & yes & \cite{emonts14, emonts15b}\\ 
MRC0156-252  & 2.016 & CO(1--0) & 60\,kpc & no & \cite{emonts14}; Nilsson et al. in prep.\\
MRC2048-272  & 2.060 & CO(1--0) & 55\,kpc & no & \cite{emonts14}; Nilsson et al. in prep.\\
MRC1138-262    & 2.161 & CO(1--0) & 30-40\,kpc & yes & \cite{emonts13}; Gullberg et al. in prep.\\
TXS0828+193  & 2.572 & CO(3--2) & 80\,kpc & no & \cite{nesvadba09}\\
MRC0943-242  & 2.923 & CO(8--7) & 80\,kpc & no & \textit{This work} \\
B3J2330+3927 & 3.086 & CO(1--0)/CO(4--3) & 30\,kpc & yes & \cite{ivison12}\\
4C41.17           & 3.798 & CO(4--3) & 13\,kpc & yes & \cite{debreuck05}\\
4C60.07            & 3.8     & CO(1--0)/CO(4--3) & 30\,kpc & yes & \cite{ivison08,greve04}\\
\hline
\hline
\end{tabular}
\caption{Overview of CO detections that are not directly associated with the radio galaxy.
$a$: Separation between the AGN and the CO detection.
}
\label{table:overview}
\end{table*}

\subsection{AGN and starbursts}

Disentangling the SED into several components suggests that the starburst
activity in H$z$RGs is driven by the interaction of two (or more) gas-rich
systems.  Major mergers have been invoked to explain the high SFRs seen
in submm galaxies \citep[SMGs; e.g.][]{engel10}, though others argue that
they may rather represent the top end of a ``main sequence'' of star-forming
galaxies \citep[e.g.][]{michalowski12}. The similar morphologies
and stellar masses of \HzRG s and SMGs suggest they may be related,
potentially through an evolutionary sequence, a high-$z$ extension
of the local relation between QSOs and ULIRGs \citep{sanders88}. Both
classes reside in parent halos of similar mass \citep{hickox12}, but
direct observations of objects in transition between SMGs and QSOs
remain restricted to a few examples \citep[e.g.][]{simpson12}. This
is where detailed observational studies of type 2 AGN like \HzRG s can
play an important role, as their stellar masses, unlike type 1 AGN's,
can be accurately determined \citep{seymour07, debreuck10}.

In MRC0943-242, we actually observe both processes: in \Ygg, we observe
an AGN host galaxy with a moderate SFR, while the companion galaxy
\Thor\ is surrounded by two additional star-forming companions, \Odin\
and \Freja.

\subsection{\Loke\ is not so atypical around radio galaxies}

Our detection of CO(8--7) emission in \Loke, which is not co-incident with 
any dust or stellar emission, and is not in the direction of the axis of the radio emission, is puzzling. 
It is not the first time CO line emission has been detected around a \HzRG\
without any counterpart at other wavelengths. Table \ref{table:overview}
lists ten \HzRG\ with CO observations where the CO emission is offset
from the AGN compared to the radio sources.

\citet{nesvadba09} found CO(3--2) emission 80\,kpc from the core of the
$z=2.6$ \HzRG\ TXS0828+193. However, the CO(3--2) emission in TXS0828+193
is aligned with the radio jet, which \citet{nesvadba09} suggest could be
triggering the collapse and excitation of the gas. The offset of CO
emission from the position of the \HzRG\ is seen in other sources as well
(see Table~\ref{table:overview}) such as, e.g. 4C60.07 \citep{ivison08},
MRC0114-211, MRC0156-252 and MRC2048-272 \citep{emonts14}. \cite{ivison08}
detect CO(4--3) emission for the \HzRG \ at $z=3.8$ at two positions:
at the location of the AGN core and 7\arcsec\ SW of the AGN. The latter
component is also detected in CO(1--0) emission \citep{greve04}. Based
on the submm observations of 4C60.07, \cite{ivison08} point out a
complication when calculating dynamical masses using the extended CO
emission, due to the misalignment of the black hole with the CO emission
and dust continuum.  \citealt{emonts14} discovered CO(1--0) emission
in three \HzRG s that were significantly offset from the AGN and which
have 4.5-9.2$\times$10$^{10}$\,$M_{\odot}$ of molecular gas, but again
aligned with the radio jet. They discuss jet-induced star formation. Two
molecular components traced by CO(4--3) for 4C41.17 were detected by
\citealt{debreuck05}. These two components are also aligned with
the radio jet, but are co-spatial with the \Lya\ emission.

The submm and optical observations of MRC 0943-242, adds to this
complicated situation, with its many components, only two of which are
traced by CO emission.

\section{Are we seeing a multiphase accretion flow?}\label{sec:speculation}

Numerical simulations suggest that galaxies acquire much of their gas
through accretion flows generated by the growth of the cosmic web of
dark matter \citep[e.g.][]{agertz11,danovich15}.  Flows of gas develop
over cosmological distances and time scales, ultimately penetrating into
halos and feeding baryons onto galaxies. Because the flows are associated
with the growth of dark matter structures, galaxies are expected to be
embedded in accretion flows.  Could we be witnessing such accretion in
our combined data sets?

Progressing inwards in radius towards the host galaxy of MRC 0943-242
over many 10s of kpc from \Loke, through \Odin, \Thor, \Freja, \Bifrost\
to the redshifted absorption component 3 seen against the inner \Lya\
emission, we see little change in the velocity over this inward journey.
In the most distant emission, i.e. \Loke, the velocity offset relative
to the AGN emission is small.  This structure appears dynamically cold,
having low dispersion in all of its spectroscopic features, implying
that it is near to the plane of the sky. Aligned with this general
structure is a group of  (merging?) vigorous star-forming galaxies
associated with the emission regions, \Freja, \Thor\ and \Odin.
If the coincident \Lya\ emission from \Bifrost\ is representative of
their projected velocity, these galaxies are embedded within this overall
(in projection) linear structure. On the opposite end of the structure,
we find the AGN host galaxy, \Ygg, which has a rather mundane rest-frame
optical morphology, suggesting a heavily obscured galaxy, but not a
merger \citep[see Fig. 9 of][]{pentericci01}. Interestingly, the extended
emission line halo does not extend on the other side of the AGN.

Admittedly, we do not have a simple explanation for the distribution
of the various phases.  For example, why is most of the molecular gas
observed at the position of \Loke? An explanation could be
that the molecular gas is related to the strong, optically thick,
large covering fraction of \HI\ absorption component 2. However,
the velocity offsets (Fig.~\ref{fig:veldist}) suggest that it is rather
the less massive absorption component 3 that coincides with the CO(8-7) and
simply part of what may be a flow into \Ygg.

Only a combined ALMA + MUSE survey of radio galaxies like that of
MRC0943-242 (with deeper integrations than in this pilot project)
can determine if it is gas accretion that is fuelling the growth of
their galaxies and supermassive black-holes.  These combined data cubes
are our only way to obtain a complete picture of various phases which
emit strongly over a wide range of wavelengths. This will allow us to
study the nature of the halo gas that results from this complex
interaction. By the sensitivity of both ALMA and MUSE on the VLT are already
offering tantalising clues on how the gas in this system is being acquired
and modified by the interaction with the AGN and galaxies in this complex
system.

\section{Conclusions}\label{sec:con}

Combining the high sensitivity of ALMA and MUSE, even with very modest
integration times, has allowed us to gain insights into the environments
which explain many properties of the evolutionary path of \HzRG s.
Surprisingly, in our pilot study of the $z=2.9$ radio galaxy MRC0943-242,
we apparently find that most of the star formation is not associated
with the radio galaxy itself, but is in a companion set of
galaxies which exhibits a complex pattern of dust emission
which we associate with star formation.  We say apparently, because
the morphology of the dust emission is peculiarly distributed in three
distinct sources but roughly within the diffuse stellar continuum emission
of the source.

The MUSE data show that the continuum emitting source is embedded in
a \Lya\ emitting linear filament stretching from beyond the AGN to the
circum-nuclear emission of the AGN.  This structure of dust and \Lya\
emission aligns well in projection with a region of CO(8--7) and CO(1--0)
emitting gas (at lower significance) about 90\,kpc in projection from
the AGN.

The AGN apparently ionises a large region which is asymmetric, lying
mostly on the western side of the source. This is likely due to the gas
distribution as it tantalisingly lies in the direction where the dust,
CO line emission, and the filamentary \Lya\ emission are found.

The relationship between the \Lya\ emission and absorption and
the associate \CIV\ absorption, paint a fascinating picture of the
distribution of the warm ionised medium. There is a component of \Lya\
absorbing gas which reaches zero intensity throughout the \Lya\ emission
detected across MRC0943-242 in the MUSE data.  It clearly has a unity
covering fraction and is highly optically thick.  Since the gas is
seen as an absorption feature throughout the \Lya\ emitting filament
(\Bifrost) this thick, unity covering fraction absorber is extended,
and visible in absorption out to $>60$\,kpc. Because of its velocity and
optical thickness, it literally absorbs all of the extended \Lya\
emission in the cube.

\appendix
\section{Line ratio diagrams from \citealt{debreuck00}}

This appendix presents a range of emission line ratio diagrams that
are directly derived from the MUSE data, compared photo-ionization and
shock ionization models (Fig.~13 from \citealt{debreuck00}).

\begin{figure*}
\includegraphics[trim=1.5cm 5.1cm 1cm 5.5cm, clip=true, scale=0.92]{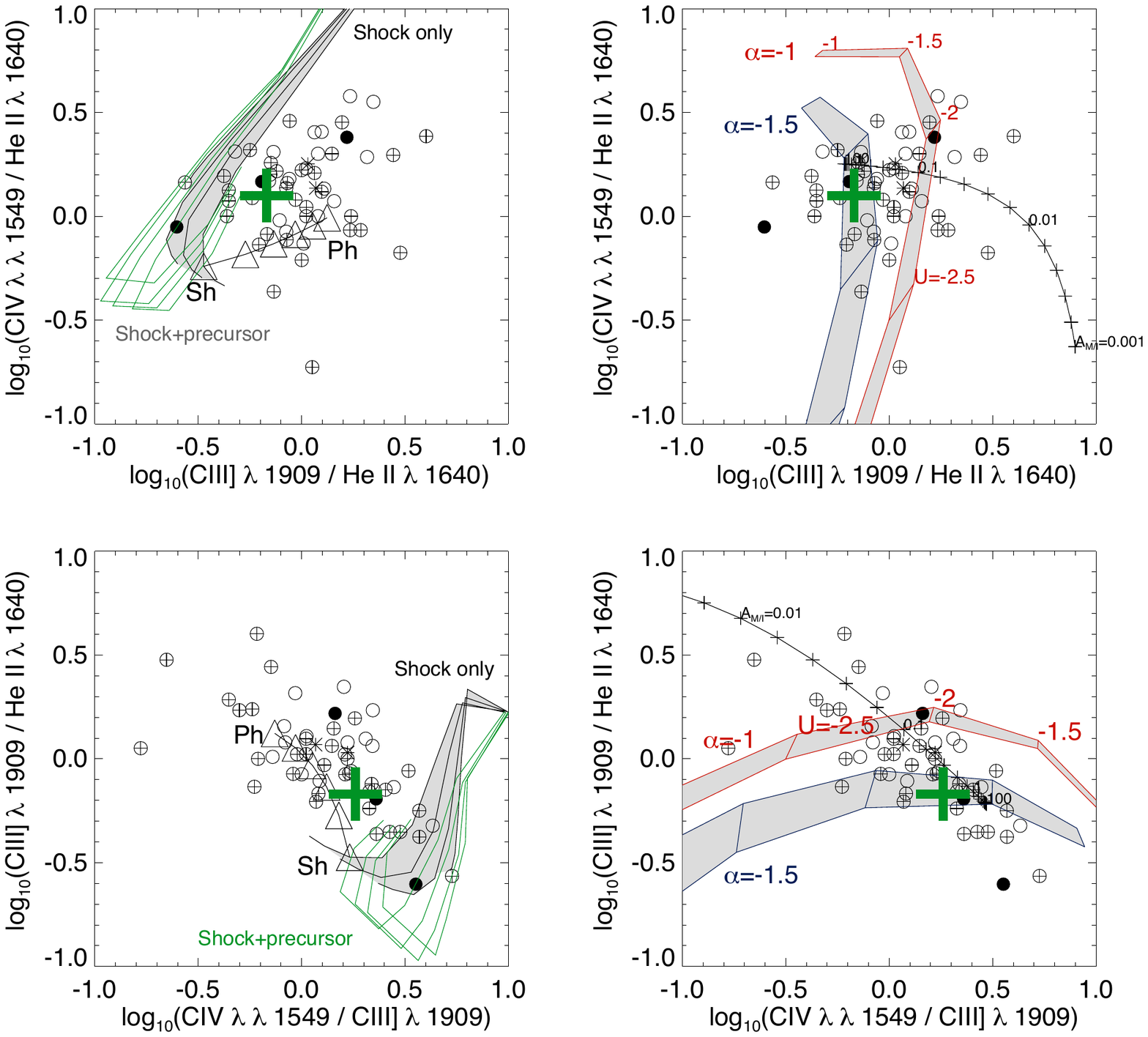}
\\
\includegraphics[trim=0.7cm 0cm 1cm 20cm, clip=true, scale=1]{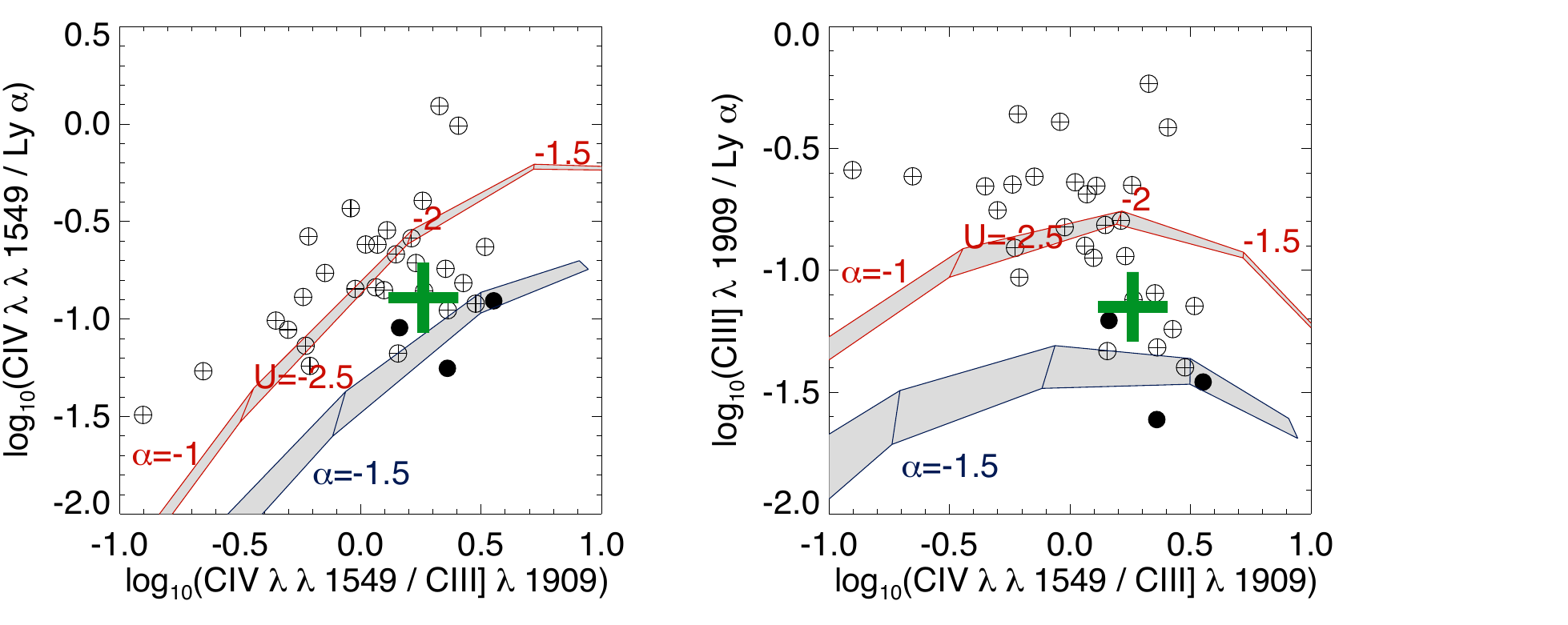}
\caption{\small  Line ratio diagnostic diagrams involving \Lya, \CIV,
\HeII, \CIII\ and \CII\ (adapted from  \citealt{debreuck00}, their
Fig.~13). The thick green cross shows the flux line ratios observed for
\Ygg, while the circles show other HzRGs from the literature. Also shown
are photo-ionization and shock models to illustrate that the observed line
ratios in \Ygg\ are dominated by photo-ionization with a contribution of
up to $\sim$30\% by shocks \citep[see][for more details]{debreuck00}. }
\label{fig:lineratios}
\end{figure*}

\begin{acknowledgements}
We thank the anonymous referee for her/his very thorough reading of our manuscript, and suggestions that substantially improved our paper.
This publication uses data taken from the MUSE commissioning run
060.A-9100. All of the MUSE data used in this paper are available through
the ESO science archive. This paper makes use of the following ALMA data:
ADS/JAO.ALMA\#2012.1.00039.S. ALMA is a partnership of ESO (representing
its member states), NSF (USA) and NINS (Japan), together with NRC (Canada)
and NSC and ASIAA (Taiwan), in cooperation with the Republic of Chile. The
Joint ALMA Observatory is operated by ESO, AUI/NRAO and NAOJ. The work
of DS was carried out at Jet Propulsion Laboratory, California Institute
of Technology, under a contract with NASA. BE acknowledges funding through the European Union FP7 IEF grant nr. 624351. MDL acknowledges the support
from the ESO visitors program and especially would like to thank Mario
van den Ancker for his help and Eric Emsellem for interesting scientific
discussions. Nick Seymour is the recipient of an ARC Future Fellowship.
\end{acknowledgements}


\bibliographystyle{aa}
\bibliography{bibtex/HzRG_bib}

\end{document}